\documentclass[]{emulateapj}
\usepackage{ifthen}
\usepackage{amsmath}
\newboolean{astroph}
\setboolean{astroph}{true}

\newcommand{\Msun}{\ensuremath{M_\odot}}

\newcommand{\Lsun}{\ensuremath{L_\odot}}
\newcommand{\Lbcg}{\ensuremath{L_{\mathrm{BCG}}}}
\newcommand{\Mcl}{\ensuremath{M_{\mathrm{cluster}}}}
\newcommand{\MsL}{\ensuremath{M_*/L}}
\newcommand{\ML}{\ensuremath{M/L}}
\newcommand{\Reff}{\ensuremath{{R_{\mathrm{e}}}}}
\newcommand{\G}{\ensuremath{\mathrm{G}}}
\newcommand{\dd}{\ensuremath{\mathrm{d}}}

\newcommand{\kmps}{\ensuremath{\mathrm{km\ s}^{-1}}}

\newcommand{\LCDM}{\ensuremath{\Lambda\mathrm{CDM}}}
\newcommand{\h}{\ensuremath{\mathrm{h}_{70}}}

\newcommand{\citebut}[2]{(\citealp{#1}; \citealp[but see][]{#2})}

\newcommand{\Eext}{\ensuremath{E_{\mathrm{ext}}}}
\newcommand{\Eint}{\ensuremath{E_{\mathrm{int}}}}
\newcommand{\Mint}{\ensuremath{M_{\mathrm{int}}}}
\newcommand{\Mext}{\ensuremath{M_{\mathrm{ext}}}}
\newcommand{\Mdm}{\ensuremath{M_{\mathrm{DM}}}}
\newcommand{\Mhalo}{\ensuremath{M_{\mathrm{halo}}}}
\newcommand{\Mbh}{\ensuremath{M_{\mathrm{BH}}}}
\newcommand{\Ri}{\ensuremath{R_{\mathrm{int}}}}
\newcommand{\Rf}{\ensuremath{R_{\mathrm{fin}}}}

\newcommand{\figcartoon}{
  \begin{figure}[htbp]
    \centering
    \plotone{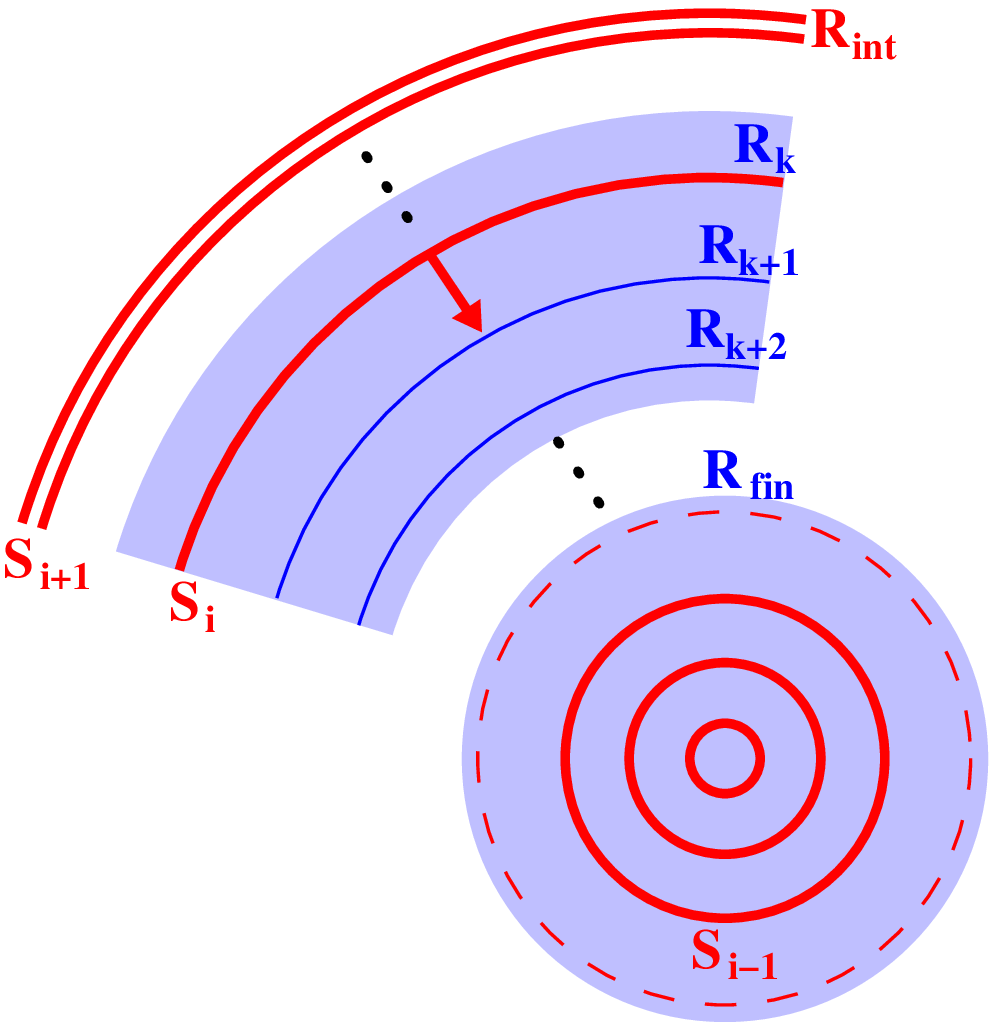}
    \caption
    {A cartoon depiction of the process by which we move stellar
      shells in to form a galaxy. The thin stellar shells are shown in red and
      labeled by $S_i$. The dark matter background is shown in blue
      and discretized into layers of uniform density and width $R_k -
      R_{k+1}$. The stellar shell $S_i$ is being moved from $R_k$ to
      $R_{k+1}$ as shown by the arrow. }
    \label{fig:cartoon}
  \end{figure}
}

\newcommand{\figvcirc}{
  \begin{figure}[htbp]
    \centering
    \plotone{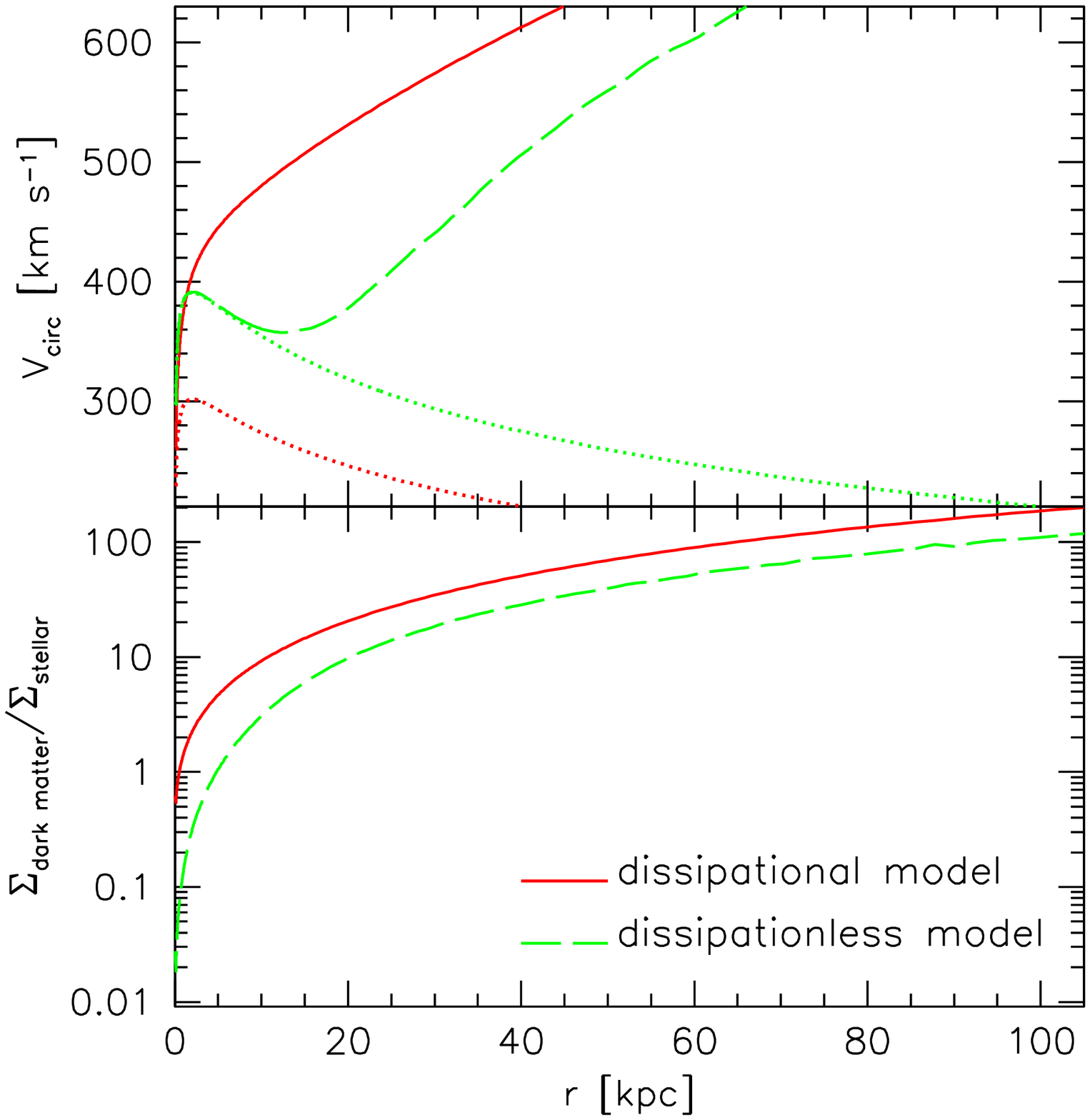}
    \caption
    {The circular velocity (top) and projected dark matter to stellar
      density ratio (bottom) for an extremely massive $L=6.0\ 
      L_*$ ($K$-band) galaxy in a $3.84 \times 10^{14}\ \Msun$
      halo. The Sersic index is $n=6.80$ and the half light radius is
      45.79 kpc. The stellar mass-to light ratios of the 
      models are set to the values shown in the lower panel of Figure
      \ref{fig:Lsigma}, such that the two models have the same central
      velocity dispersion ($\sigma_0 \approx 270 \ \kmps$). The solid
      line is for the dissipational model. The
      dashed line is for the dissipationless model.  In the upper
      plot, the upper (dissipationless) and lower (dissipational)
      dotted lines are the circular velocity curves due to the stars
      alone. }
    \label{fig:vcirc}
  \end{figure}
}

\newcommand{\figLsigma}{
  \begin{figure}[btp]
    \centering
    \plotone{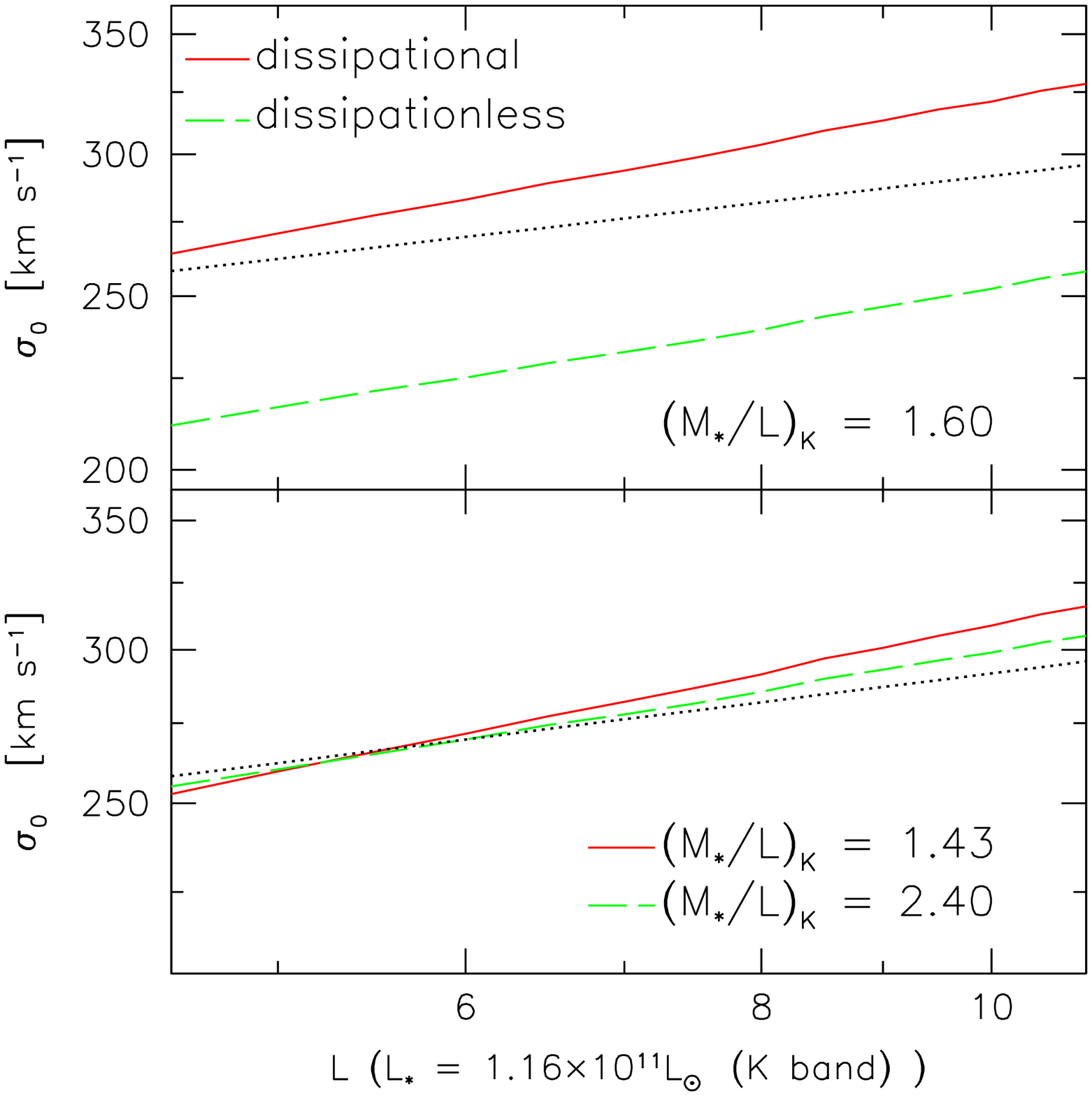}
    \caption
    {The $L$--$\sigma$ relation in the $K$-band from \citet{Lauer07}
      compared to the calculated relation for both the dissipational
      and dissipationless models. The dotted line is the relation from
      \citet{Lauer07}. The top figure shows both models at
      $(\MsL)_K = 1.60$. By adjusting the stellar
      mass-to-light ratio for each model separately, the model lines can be
      made to overlay the observed relation, as shown in the lower panel. }
    \label{fig:Lsigma}
  \end{figure}
}

\newcommand{\figlenses}{
  \begin{figure*}[tbp]
    \centering
    \includegraphics[angle=270,width=\textwidth,trim = 170 0 0 0,
    clip]{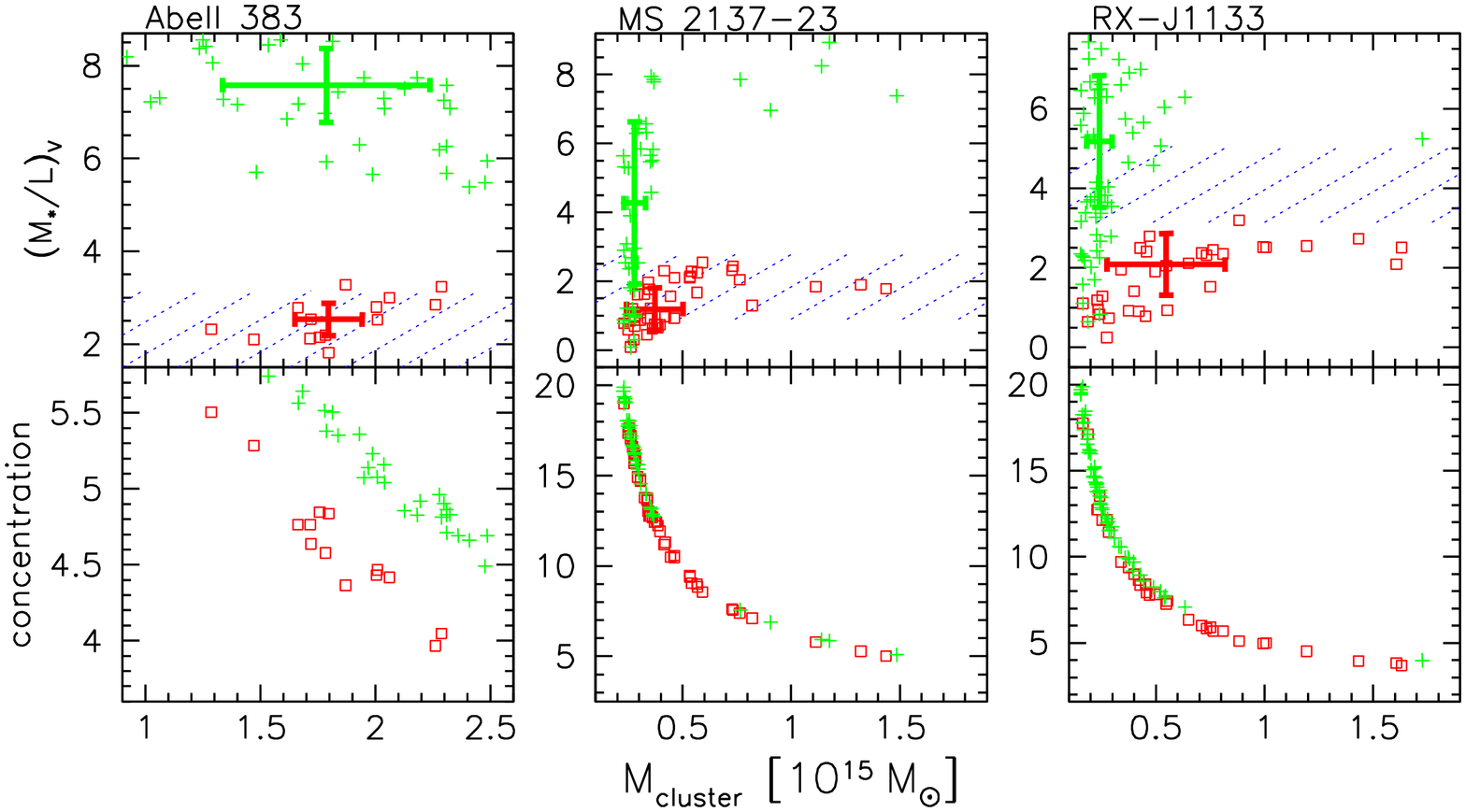} 
    \caption
    { The best-fit dissipational (squares) and dissipationless
      (crosses) models from a random distribution of input parameters:
      \MsL, \Mcl, and $c$. The points shown for Abell 383 give
      results within
      $2\sigma$ of the measured velocity dispersion, and radial and
      tangential lensing arc measurements. For the other galaxies, the
      points represent the models within $1\sigma$ of lensing and
      dynamics observations. The heavy lines  in the \Mcl-\MsL{} plot
      show the median and central 50\% of the points. The shaded
      regions show the range of expected values for \MsL{}, assuming
      passive evolution of the stellar populations Note: The plot
      for RXJ-1133 shows the $B$-band stellar mass-to-light ratio.}
      \label{fig:lenses}
  \end{figure*}
}

\newcommand{\figmsbestfit}{
  \begin{figure}[tbp]
    \centering
    \plotone{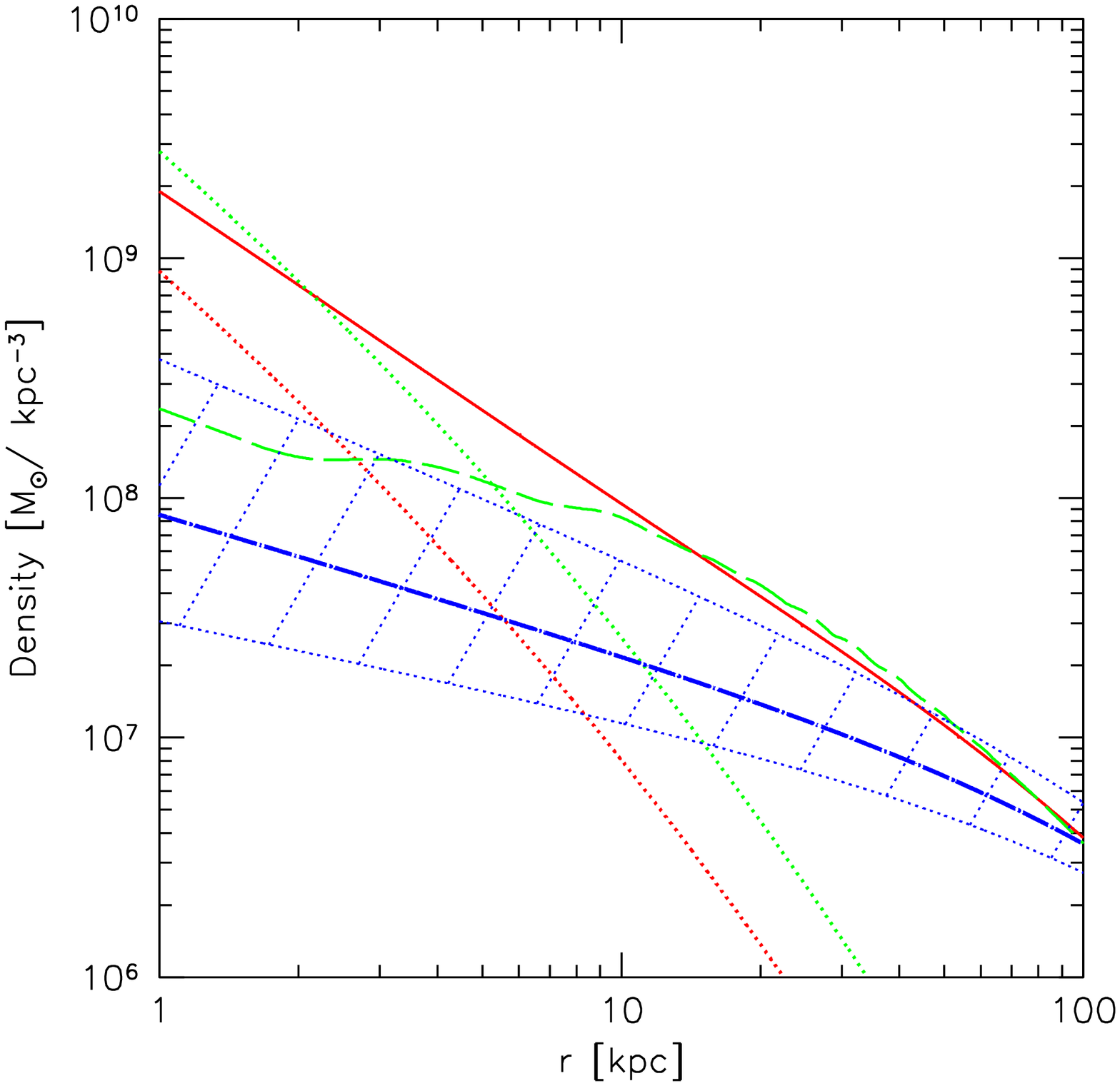}
    \caption
    {The dark matter and stellar density profiles for two best-fit
      models to the lensing observations of MS 2137-23. The solid
      curve is the dark matter and the lower dotted curve is the
      stellar component of the dissipational 
      model. The dashed curve and the upper  dotted curve
      are the dark matter and stars, respectively, of the
      dissipationless model.  The dash-dotted curve with the
      shaded region is the best-fit model for the dark matter from
      \citet{SandTreu04} 
      along with $2\sigma$ error bars. }
      \label{fig:msbestfit}
  \end{figure}
}

\newcommand{\figMratio}{
  \begin{figure}[htbp]
    \centering
    \plotone{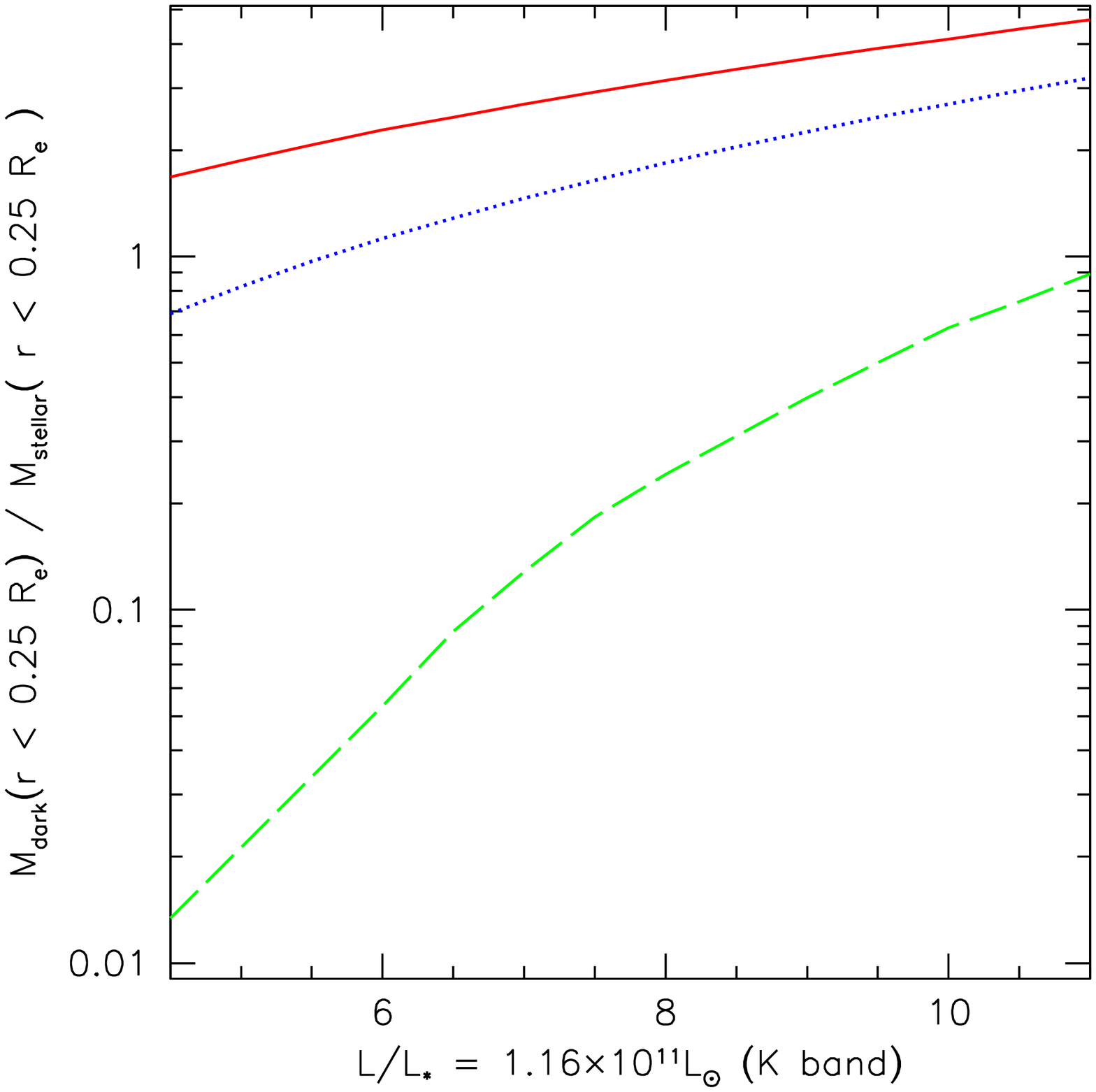}
    \caption
    {The ratio of dark to stellar matter contained within $0.25\, \Reff$
      as a function of BCG luminosity. The solid line shows the
      dissipational model, the dashed line shows the dissipationless
      model. The dotted line in the center shows the ratio of an NFW
      profile for the dark matter to a Sersic profile for the stars. }
      \label{fig:Mratio}
  \end{figure}
}

\newcommand{\figSAURON}{
  \begin{figure}[htbp]
    \centering
    \plotone{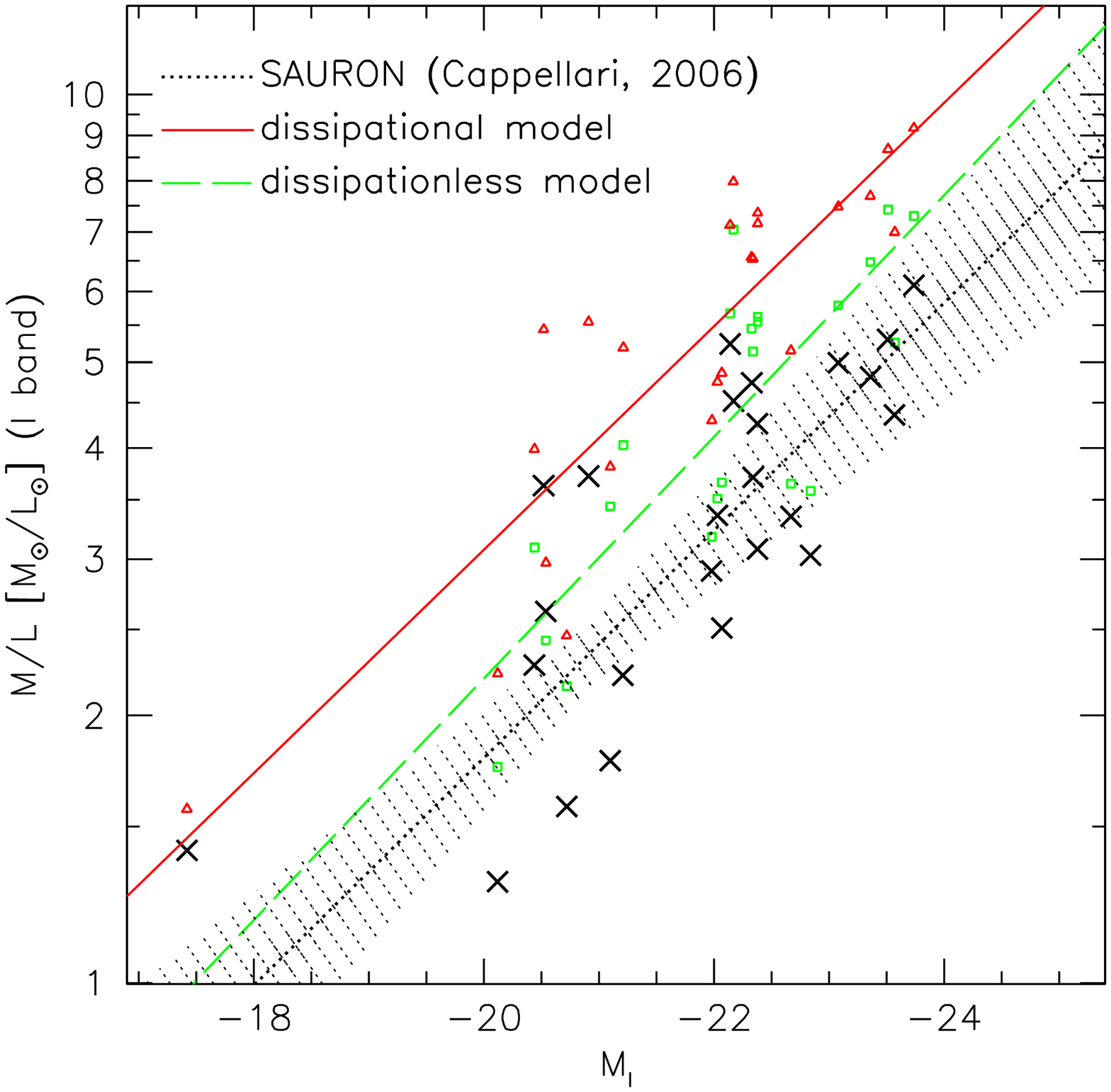}
    \caption
    {The trend in \ML{} with galaxy luminoisty observed in the SAURON
      data \citep{Cappellari06}. The \MsL{} values for these galaxies
      are all below 3.4. The `$\times$'-symbols denote the SAURON
      data. The dotted line and surrounding shaded region show the
      best fit to the SAURON data. The triangles (squares) show the
      \ML{} calculated for the dissipational (dissipationless)
      models for each SAURON galaxy. The solid and dashed lines show
      the best fit for the dissipational and dissipationless models,
      respectively. }
      \label{fig:sauron}
  \end{figure}
}

\newcommand{\figPNe}{
  \begin{figure}[tbp]
    \centering
    \plotone{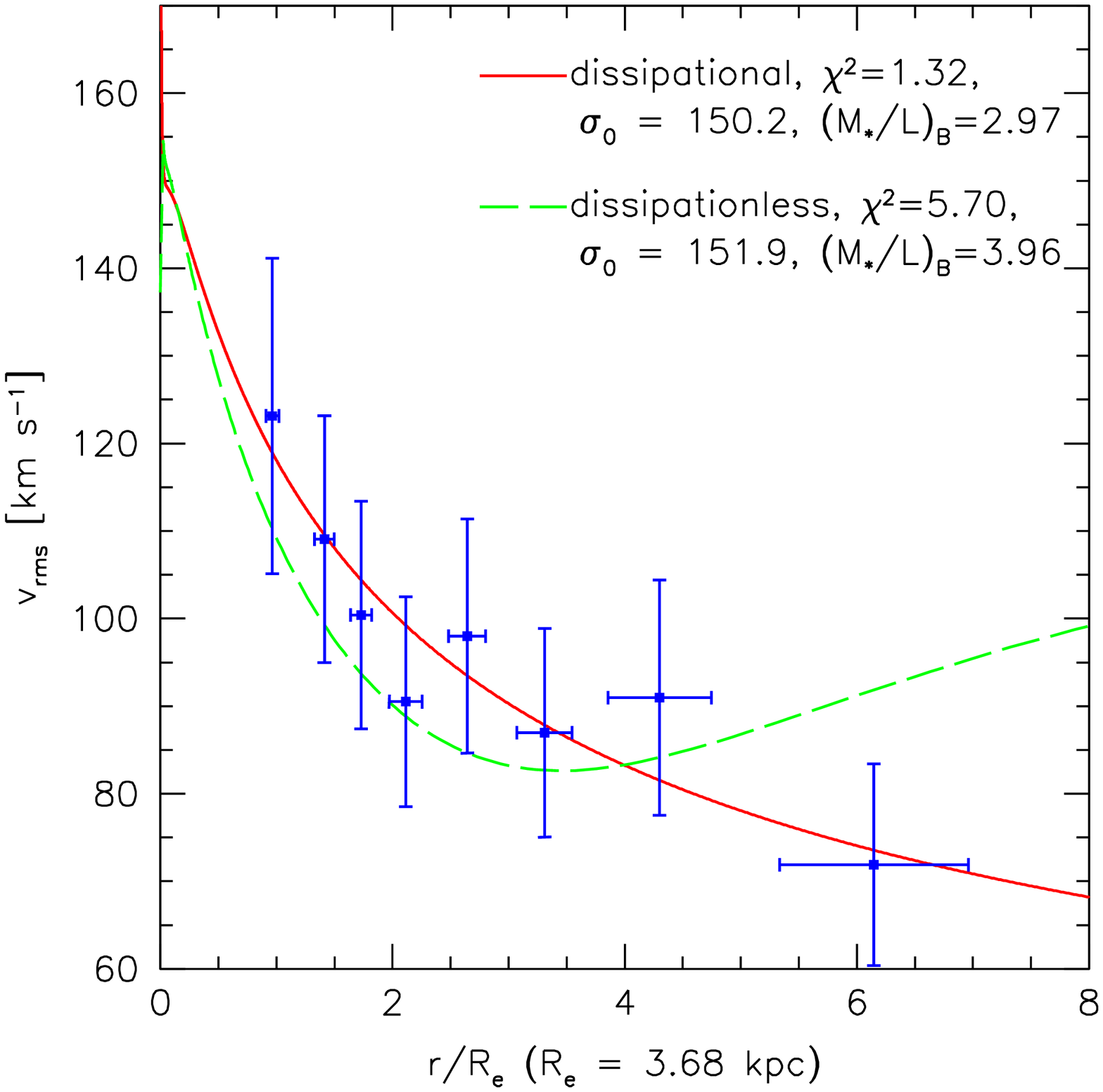}
    \caption
    {The velocity dispersion of NGC 4494 from \citet{Napolitano09}
      measured using planetary nebulae.  The two lines show the
      dissipational and dissipationless galaxy formation
      models. The central velocity
      dispersion of NGC 4494 is taken to be
      $150.2\pm3.7\ \kmps$ \citep{Paturel03}. 
    }
    \label{fig:PNe}
  \end{figure}
}

\newcommand{\figPNeDMdensity}{
  \begin{figure}[bhp]
    \centering
    \plotone{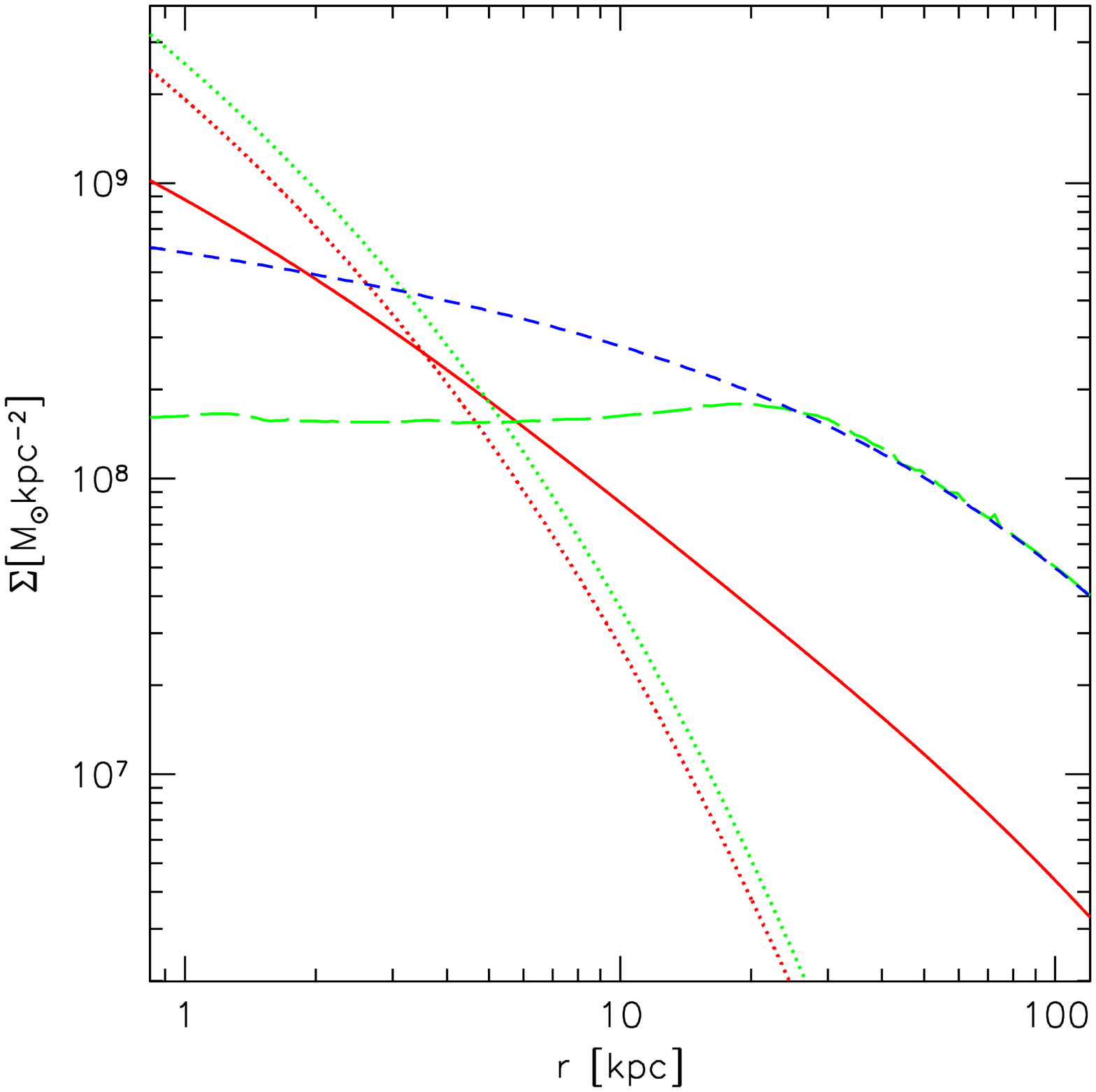}
    \caption
    {The projected dark matter density profile for the best-fit models to NGC
      4494. The solid line is for the dissipationless model. The
      long-dashed line is the dissipationless model. The short-dashed
      line 
      is the best fit NFW. The two dotted lines are the stellar
      profiles for the dissipationless (upper) and dissipational
      (lower) models. 
    }
    \label{fig:PNeDMdensity}
  \end{figure}
}

\newcommand{\tablemodels}{
\begin{deluxetable}{lcc}
\tabletypesize{\scriptsize}
\tablecaption{Input model parameters for strong lensing BCG and
  cluster: MS 2137-23\label{table:models}}
\tablewidth{0pt}
\tablehead{
\colhead{Parameter} & \colhead{Dissipational} &
\colhead{Dissipationless} }

\startdata
\cutinhead{Free (fitted) model parameters}
$\log\Mcl/\Msun$ & 14.75 & 14.55 \\
concentration ($c$) & 8.99 & 13.27 \\
$(\MsL)_V$ & 1.67 & 5.47 \\
\cutinhead{Observed model parameters\tablenotemark{a}}
$\log(\Lbcg{}/\Lsun{})_V$ & $11.5$ & $11.5$ \\
$n_{\mathrm{Sersic}}$ & 4.0 & 4.0 \\
$\Reff\ [\mathrm{kpc}]$ & $23.03$ & $23.03$ \\
$\sigma_{0}\  [\kmps]$\tablenotemark{b}& 317.33 & 321.90 \\
\enddata
\tablenotetext{a}{Observations of Sersic fit from \citet{SandTreu04}.}
\tablenotetext{b}{Observed $\sigma_0 =
  319\pm26\ \kmps$ \citep{SandTreu04}. The numbers given 
  are from the best-fit models.} 
\end{deluxetable}
}
\ifthenelse{\boolean{astroph}}{
  \newcommand{\middfig}[1]{{#1}}
  \newcommand{\tailfig}[1]{{}}
  
}{
  \newcommand{\tailfig}[1]{{#1}}
  \newcommand{\middfig}[1]{{}} 
}

\newcommand{\myemail}{clackner@astro.princeton.edu}

\shorttitle{Dark Matter in Centers of Galaxies}
\shortauthors{Lackner \& Ostriker}

\begin{document}

\submitted{Accepted to ApJ}

\title{Dissipational versus Dissipationless Galaxy Formation and the Dark
  Matter Content of Galaxies}

\author{C.N. Lackner and J.P. Ostriker}
\affil{Department of Astrophysical Sciences, Princeton University,
  Princeton, NJ 08544}
\email{\myemail}

\begin{abstract}
We examine two extreme models for the build-up of the stellar
component of luminous elliptical galaxies. In one case, we assume the
build-up of stars is dissipational, with centrally accreted gas
radiating away its orbital and thermal energy; the dark matter
halo will undergo adiabatic contraction and the central dark matter
density profile will steepen. For the second model, we assume the
central galaxy is assembled by a series of dissipationless mergers of
stellar clumps that have formed far from the nascent galaxy. In
order to be accreted, these clumps lose their orbital energy to the
dark matter halo via dynamical friction, thereby heating the central dark
matter and smoothing the dark matter density cusp. The central dark
matter density profiles differ drastically between these models. For
the isolated elliptical galaxy, NGC 4494, the central dark matter densities
follow the power-laws $r^{-0.2}$ and $r^{-1.7}$ for the dissipational
and dissipationless 
models, respectively. By matching the dissipational and dissipationless 
models to observations of the stellar component of elliptical
galaxies, we examine the relative contributions of dissipational 
and dissipationless mergers to the formation of elliptical galaxies and
look for observational tests that will distinguish between these
models. Comparisons to strong lensing brightest cluster galaxies yield
median $(\MsL)_B$
ratios of $2.1\pm0.8$ and $5.2\pm1.7$ at $z\approx0.39$ for the
dissipational and 
dissipationless models, respectively. For NGC 4494, the best-fit
dissipational and dissipationless models have $(\MsL)_B=2.97$ and
$3.96$. Comparisons to expected stellar mass-to-light ratios from
passive evolution and population syntheses appear to rule out a
purely dissipational formation mechanism for the central stellar
regions of giant elliptical galaxies.
\end{abstract}

\keywords{ galaxies: formation---galaxies: elliptical and lenticular,
  cD---dark matter }

%
\section{Introduction}
\label{sec:intro}

How is the stellar content of  galaxies assembled? Two extreme,
contrasting models for the assembly of stars in galaxies
have been considered over the years with no conclusive evidence as
yet as to which mode dominates in which systems at which times. At one
extreme, we can treat the process as totally dissipational with
regard to energy; gas flows in from the virial radius, radiating away
the kinetic and thermal energy it acquires while descending into the
deep potential well of the dark matter halo. Once in place, the gas is
transformed into stars `in situ'--in approximately the regions in
which we see stars in fully-formed galaxies today. The other extreme
model postulates that stars are formed in smaller stellar systems far
outside the 
effective radius of the ultimate galaxy. From there, they lose orbital
energy via dynamical friction and lose their potential energy only
by heating or expelling other matter. These stars could be considered
`accreted,' whether by minor or major mergers. The balance between
these two processes of galaxy formation (both of which surely occur) is unknown;
determining that balance should help us to unravel some
apparent paradoxes in the standard \LCDM{} model of cosmology.

Although the \LCDM{} model of cosmology has been very successful explaining
large scale observations, such as the cosmic microwave background and
the large scale structure of galaxies, it has not enjoyed as much
success on smaller, galactic scales.  One notable discrepancy between
simply-modeled \LCDM{} predictions and observations is the difference
between the predicted and observed dark matter content in the centers
of galaxies. Many N-body simulations of dark matter particles have
been performed, and they show that cold dark matter will
collapse into self-similar halos with a central density cusp
\citebut{Fukushige97, NFW97, Moore99, Diemand05}{Springel08}. The
steepness of the 
cusp may vary from $r^{-1}$ in the case of the NFW profile \citep{NFW97}
to as steep as $r^{-1.5}$ \citep{Moore99}. Other studies
\citep{Subramanian00, Ricotti03} have shown a
range of central slope indices, $1<\alpha<1.5$ for $r^{-\alpha}$,
depending on time, mass, 
and environment.

Cold dark matter simulations uniformly predict that dark matter halos
should be universally cuspy, but observations have yet unambiguously to
find these cusps.  Indeed, observations of gravitational lensing, stellar
velocity dispersions, and gas dynamics suggest inner dark matter
density profiles are cored ($r^\alpha$, $\alpha > -1$) instead of
cuspy \citebut{FloresPrimack94, Romanowsky03, Swaters03, Gentile04,
  SandTreu04, Simon05, Cappellari06, Gilmore07, Gentile07, deBlok08,
  Oh08, Napolitano09}{Rhee04, Spekkens05, Valenzuela07}. For the Milky 
Way, it has been shown that within the errors of the microlensing
observations, stars alone can more than account for the total mass
density of the galaxy, leaving little room 
for a dark matter component in the center \citep{Binney01}.

Acceptance of both the dark matter simulations and the observations of
dark matter 
density profiles leaves us with an apparent discrepancy. One way to solve this
discrepancy is to abandon \LCDM{}. For example, the self-interactions
of warm dark matter will smooth central density cusps
\citep{SpergelSteinhardt00, Bode01}. Another solution to 
this discrepancy may lie in a misunderstanding of galaxy formation and
the assembly of stellar material at the center of dark matter
halos. The addition of baryons to dark matter halos can have a profound
effect on the dark matter profile of a galaxy. One such effect is the
adiabatic contraction of dark matter by the slow addition of baryons to
the center of the potential well \citep{Blumenthal86}. Adiabatic
contraction has been observed in simulations with both dark matter and
baryons \citep{Navarro91, Navarro94, Jesseit02, Abadi03, Gnedin04}. This process
conserves the adiabatic invariants of the dark matter orbits. With
regard to energy, however, it is a
dissipational process, as the orbital energy of the infalling baryonic
material is radiated away and lost from the system. Because the dark
matter density increases in the  center of the galaxy under
adiabatic contraction, the discrepancy between simulations and
observations is only made worse, in the sense that the inner slope
would be steeper than the NFW slope. For example, if the ultimate mass
profile is ``isothermal,'' ($\rho_{\mathrm{tot}}(r)\propto r^{-2}$), then a dark
matter profile with an initial slope index  $1<\alpha<1.5$,
($r^{-\alpha}$), would have a post-adiabatic-contraction index of
$1.67<\alpha<1.8$.  

Processes which lower the dark matter density in the central regions
of galaxies have also been explored. These processes are dissipationless; energy
from stellar baryons is transferred to the dark matter, heating it and lowering
the central dark matter density.  Such processes include interactions
of the dark matter with a
stellar bar \citep{Weinberg02, HolleyBocklemann05, McMillanDehnen05},
baryon energy feedback from AGN\citep{Peirani08}, decay of binary
black hole orbits after galaxies merge \citep{Milosavljevic01},
scattering of dark matter particles by gravitational heating from
infalling subhalos \citep{MaBoylanKolchin04}, and dynamical friction
of stellar/dark matter clumps against the smooth background dark
matter halo \citep{El-Zant01, El-Zant04, Tonini06, Romano-Diaz08,
  RomanoDiaz09, Jardel09}. In order to be successful, these methods
must retain the strong central stellar concentration observed in large
galaxies.

\citet{El-Zant01} propose erasing the dark matter cusp via
dynamical friction of incoming stellar/dark matter clumps against the
dark matter background.  \citet{Romano-Diaz08} tested this
hypothesis with dark matter/baryon N-body/SPH simulations and found
that the introduction of baryons can flatten the dark matter cores in
the inner 3 kpc.  \citet{Romano-Diaz08} claim that this cusp-flattening
is not seen in other baryon/DM simulations because of a low numerical
resolution and a focus on early, dissipational galaxy
formation. Recent simulations \citep{Naab07, Abadi09, Johansson09}
including both dark matter 
and baryons have shown similar departures from the standard adiabatic
contraction model. In their cosmological simulations for building up 
elliptical galaxies, 
\citet{Johansson09} note that their results show reasonably low dark
matter fractions in the inner $10\ \mathrm{kpc}$ and that the assembly
of their elliptical galaxies at late times is dominated by accretion of stellar
lumps, not gas, which as noted above, tend to reduce the dark matter
concentration.  Similarly,
accretion at late times has been invoked to
erase the dark matter cusp formed by early adiabatic contraction, bringing
dark matter halos to a universal (e.g. NFW) profile
\citep{LoebPeebles03, Gao04}.  

Dynamical friction of stellar lumps
against the dark  
matter halo represents a dissipationless method of stellar build-up in
which the orbital energy of the infalling stars is transferred to the
dark matter, thereby heating it and driving it out of the central
parts of the galaxies. This is in direct contrast to the case
of adiabatic contraction, which arises from a dissipational build-up of
stellar material.  Although both processes undoubtedly take place, the
fully dissipational (adiabatic contraction) and fully
dissipationless (dynamical friction) scenarios for galaxy
formation represent the two extrema. Since these two
processes have opposite effects on the central dark matter content of
galaxies, the balance between them will determine the present-day dark
matter density profiles in galaxies.

In this paper we explore the physics of galaxy assembly, presenting
two toy models for the two extremes of galaxy
formation constrained to produce the same observed final stellar
distributions: one model for the fully dissipational build-up of
stellar matter with star
formation occurring in-situ and one model for the fully
dissipationless build-up of stellar matter with stars added by
accretion. The models are described in \S\ref{sec:models}. We focus
on the structure of giant elliptical galaxies whose light profiles are
well-described by Sersic models. Taking the observed stellar
profile of the galaxy as given, but leaving the stellar mass-to-light
ratio, \MsL{}, as a free
parameter, we  model the assembly of the
stellar component via the two extreme formation methods. Then, by comparing
the properties of the galaxies formed by these two methods to
observations, we can begin to determine which method of assembling stars
dominates and help to resolve the discrepancies between the simulated
and observed dark 
matter density profiles.

Throughout this paper, we adopt the \LCDM{} cosmology with  $H_0 = 70\
\h\ \kmps$, $\Omega_{\mathrm{baryon}}/\Omega_{\mathrm{matter}} =
0.17$, $\Omega_{\mathrm{matter}} = 0.26$, and $\Omega_{\Lambda}=0.74$.

\section{Models}
\label{sec:models}

In this section, we describe the two toy models of galaxy
formation used in this paper. The first assumes that the orbital
energy of the in-falling baryons is deposited entirely in the dark matter halo,
while the second assumes energy is radiated away, leaving the galaxy-halo
system. 

In both models, we assume that both the dark matter and the baryonic
matter follow circular orbits. For simplicity, we also assume the initial
conditions (before the formation of the galaxy) for both the dark
matter and the baryons are NFW density profiles, $\rho \propto
[(r/r_{\mathrm{NFW}})(1+r/r_{\mathrm{NFW}})^2]^{-1}$, with the same
concentrations. The ratio of baryon to dark matter density is equal to
the universal fraction, which we assume to be
$\Omega_{\mathrm{baryon}}/\Omega_{\mathrm{dark matter}} = 0.20$. In
reality, the baryons will have a broader distribution than the dark
matter, because the baryons are coupled to the radiation
background. However, the additional thermal energy of the baryons
corresponds to a velocity less than $10\ \kmps$, which is negligible
for halos more massive than $\sim10^8\ \Msun$. In this study, we
restrict ourselves to large halos, and can safely assume the baryons
are initially distributed identically to the dark matter. We have
verified this is a valid assumption by placing all the baryons
initially outside the virial radius where they have negligible binding
energy. This does not affect any of the results in this paper,
altering the dark matter mass with $5\ \Reff$ by a few parts in $10^3$.

The final condition to which the models evolve is an elliptical galaxy
with a predetermined stellar luminosity profile in the center of a
dark matter halo. In both the dissipational and dissipationless
models, we assume that the final galaxies contain no gas; all the
baryons are accounted for in stars. This assumption is justified
because the gas mass in elliptical
galaxies today is usually only a few percent of the total baryon
mass in these galaxies \citep{Georgakakis01}. 

For both models, the dark matter and stellar matter profiles are
discretized into spherical shells such that the radius of a shell is
at least two orders of magnitude larger than its width, and each shell
is taken to be of homogeneous density.  In order to
form the galaxy, shells of stellar 
matter are moved inwards.  Each time a shell of stellar matter passes through a
shell of dark matter, the stars' orbital energy is either deposited in the
dark matter shell (dissipationless model) causing the dark matter shell to
move outwards, or the stars' orbital and thermal energy is
radiated away causing the dark matter shell's average radius to decrease as it
undergoes adiabatic contraction. Since truly spherical shells would
not suffer from dynamical  
friction, we are in fact considering a shell composed of stellar
clumps, all of which (at any given time) have the same energy and
angular momentum per unit mass, but whose angular momentum vectors are
oriented randomly, giving the shell a total angular momentum of
zero. Additionally, we assume that the stellar clumps do not contain
any dark matter, as they formed far from the center of the dark matter
halo, where the density is low. In reality, these clumps will contain
some dark matter, much of which will be stripped before the stellar
clump merges with the galaxy. Any remaining dark matter will undergo
dynamical friction against the background dark matter. 
Because the system is spherically symmetric, a given dark matter shell
is only affected by matter that crosses through it, but not by the
rearrangement of matter interior or exterior to it. Therefore a
procedure that describes a single shell crossing can be repeated many
times for many shells, until the desired galaxy has been assembled.

\subsection{Dissipationless Model: Dynamical Friction}
\label{ssec:modelDF}

In the dissipationless model, all the orbital energy lost by the
stars in forming the galaxy is deposited in the dark matter halo.  In
this case, the galaxy is built up by `dry,' dissipationless
mergers.

We imagine constructing the Sersic profile of the stars in a shell by
shell fashion, sequentially moving each shell of stars from a large
radius to its final position, in such a way that two stellar shells
never cross. Therefore, the problem can first be idealized as moving
one infinitely thin stellar shell from a large radius $R_{\mathrm{i}}$
to a smaller, final radius $\Rf$. As the stellar shell moves, it
passes through a dark matter distribution with a mass distribution
$M_{\mathrm{dark}}(r)$, to which the stellar shell gives its orbital
energy. The dark matter can be subdivided into a series of mass shells
of thickness $\Delta R_k$, such that $\sum_k \Delta R_k = \Ri -
\Rf$. Therefore, we only need to compute the effects of one stellar
shell passing through one dark matter shell uniformly distributed
between $R_{k}$ and $R_{k+1}$, with thickness $\Delta R_k$, and then
execute a double sum over the dark matter shells and all the stellar
shells. A depiction of this process is shown in Figure
\ref{fig:cartoon}. In the figure, the sum over dark matter shells is
indexed by $k$, while the sum over the stellar shells is indexed by
$i$.  

\middfig{\figcartoon}

The total energy lost by a single stellar shell as it moves from $R_k$ to
$R_{k+1}$ can be calculated by taking the difference between the
energy change of the stellar shell as it moves from a large initial
radius, $\Ri$, with zero potential and zero kinetic energy, to $R_{k+1}$
and the energy change as it moves from $\Ri$ to $R_{k}$. The changes in 
kinetic and potential
energy per unit mass of the stellar shell as it moves from $\Ri$ to $R_k$ are
\begin{eqnarray}
\label{eq:deltaEstars}
\Delta T_k &=& \frac{1}{2}\frac{\G M(R_k)}{R_k}\,,\qquad\mathrm{and} \\
\Delta W_k &=& -\frac{\G M(R_k)}{R_k} - \int_{R_k}^{\Ri}\frac{\G
  \, \dd M_{\mathrm{dark}}(r)}{r} \, ,
\end{eqnarray}
where $M(r)$ includes both the dark matter and any stellar matter
interior to $r$. By assumption, there is no stellar matter between
$\Ri$ and $R_k$, so the integral in the potential energy depends only on
$M_{\mathrm{dark}}$. The change
in total energy of the stellar shell as it moves from $R_k$ to
$R_{k+1}$ is thus 
\begin{eqnarray}
\label{eq:deltaEdeltaRk}
\Delta E_k &=& \frac{\G M_{\mathrm{dark}}(R_k)}{2 R_k} - \frac{\G
  M_{\mathrm{dark}}(R_{k+1})}{2 R_{k+1}} + \\\nonumber
&&\, \frac{\G
M_{\mathrm{star}}(R_k)}{2}\left(\frac{1}{R_{k+1}}-\frac{1}{R_k}\right) +
\\\nonumber
&&\, \int_{R_k}^{R_{k+1}}  \frac{\G \,\dd M_{\mathrm{dark}}(r)}{r} \, .
\end{eqnarray}

The stellar shell is made of up of small lumps of stars
that will undergo dynamical friction against the dark matter
background as they move; the energy lost by the stars as
they move from $R_k$ to $R_{k+1}$ must be deposited in the dark
matter. We assume that the exchange of energy between stars and dark
matter is a local process. Therefore, all the energy of the stars is
deposited in the dark matter initially orbiting between $R_k$ and
$R_{k+1}$. As the dark matter gains energy, it's orbit will expand,
and the uniform density dark matter shell between $R_k$ and $R_{k+1}$
will become wider. Since we are assuming the energy exchange is
local, the dark matter orbiting at the inner radius ($R_{k+1}$) will
not move. In 
reality, this dark matter is affected by both adiabatic contraction,
since the stars add mass interior to $R_{k+1}$, and dynamical
friction, since the 
energy exchange is not purely local; this layer will move, but the
movement will be second order in $\Delta R_k = R_{k} - R_{k+1}$. Therefore,
by conserving energy the only quantity which changes is thickness of
the dark matter shell the stars have moved through and consequently
its mean radius.

In order to calculate the new dark matter layer thickness, we repeat
the procedure above, but this time also account for the energy of the
dark matter shell both before and after the stars move through it. As
above, we assume a spherical dark matter shell of mass $\Mdm$ and
uniform density, with an inner radius of $R$ (corresponding to
$R_{k+1}$ above), and a thickness $\Delta R << R$. Directly exterior
to  the dark matter shell is an infinitesimally thin stellar shell of
mass $\Delta M_*$, at the radius $R+\Delta R$. The kinetic+potential energy
of the two-shell system can be broken into the self-interaction energy
of the dark matter shell, $E_{\mathrm{d}}$, the self-interaction
energy of the stellar shell, $E_{\mathrm{s}}$, and the interaction
energy of the two shells, $E_{\mathrm{ds}}$. To first order in $\Delta
R$, these are given by:
\begin{eqnarray}
\label{eq:EdB}
E_{\mathrm{d}} &=& -\frac{\G \Mdm^2}{4 R}\left(1-\frac{2}{3}\frac{\Delta
    R}{R}\right)\, ;\\
\label{eq:EsB}
E_{\mathrm{s}} &=& -\frac{\G \Delta M_*^2}{4 R}\left(1-\frac{\Delta
    R}{R}\right)\, ;\\
\label{eq:EdsB}
E_{\mathrm{ds}} &=& -\frac{\G \Delta M_* \Mdm}{2 R}\left(1-\frac{\Delta
    R}{R}\right)\, .
\end{eqnarray}
The shells are embedded in a spherically symmetric galaxy which also 
contributes to the energy. We assume that the mass distribution of the
galaxy remains fixed as the shells interact. This assumption is
exactly true for the mass interior to the shells and the mass far
outside the shells. We define $\Mint$ to be the total mass (dark
matter $+$ baryons ) interior to $R$. This mass contributes kinetic+potential energy $\Eint$ to the shells while the
mass external to $R+\Delta R$ contributes a potential energy
$\Eext$. These energies are given by
\begin{eqnarray}
\label{eq:Eint}
\Eint &=& -\frac{\G \Mint}{2R}\bigg[\Mdm\left(1-\frac{\Delta
      R}{2R}\right) +  \\\nonumber
&&\qquad\qquad\left.\Delta M_*\left(1-\frac{\Delta
      R}{R}\right)\right]\,,\ \quad\mathrm{and}\\
\label{eq:Ext}
\Eext &=& -\G\left(\Delta M_*+\Mdm\right) \int_{R+\Delta
  R}^{\infty}\frac{\dd\,M(r)}{r} \,.
\end{eqnarray}
Therefore, the total initial energy of the two shells is given by the sum of
equations (\ref{eq:EdB})-(\ref{eq:Ext}). 

We now move the stellar shell through the dark matter shell until
the stellar shell is orbiting at the radius $R$, as in the above example.
This time, however, we will include the changes in the dark matter
distribution in the energy difference. After the move, the dark 
matter shell will widen from $\Delta R$ to $\Delta R^{\prime}$.
Because $\Delta R^{\prime}$ is still small compared to $R$, the
density of the dark matter layer is still uniform after the move. The final
total energy of the stellar and dark matter shells after moving the stars
is
\begin{eqnarray}
\label{E:after}
E &=&
-\frac{G}{2R}\bigg[\frac{\Mdm^2}{2}\left(1-\frac{2}{3}\frac{\Delta
      R^{\prime}}{R}\right) + \frac{\Delta M_*^2}{2} + \\\nonumber
&&\qquad \Delta M_*\Mdm\left(1-\frac{\Delta R^{\prime}}{2R}\right) + \\\nonumber
&&\qquad \Mint\left(\Mdm\left(1-\frac{\Delta R^{\prime}}{2R}\right) +
  \Delta M_*\right)\bigg]  \\\nonumber
&& -\G \left(\Delta M_*+\Mdm\right) \int_{R+\Delta
  R}^{\infty}\frac{\dd\,M(r)}{r} + \Delta \Eext\, . 
\end{eqnarray}
If we assume that the mass initially exterior to $R+\Delta R$ is
unaffected by the dark matter expanding to $R+\Delta R^{\prime}$, the
change in external energy, $\Delta \Eext$, depends only on the mass
initially between $R+\Delta R$ and $R+\Delta R^{\prime}$, and the dark matter that moves beyond $R + \Delta R$. This
assumption is valid to first order in $\Delta R$. Since both $\Delta
R$ and $\Delta R^{\prime}$ are small compared 
to $R$, we assume that mass between $R+\Delta R$ and $R+\Delta
R^{\prime}$ is of homogeneous density and a total mass
$\Mext$. Therefore, 
\begin{equation}
\label{eq:Eextdiff}
\Delta \Eext = \frac{\G \Mdm \Mext}{4 R}\frac{\Delta R^{\prime}-\Delta
  R}{\Delta R^{\prime}}\,.
\end{equation}
Substituting this into equation \ref{E:after} and taking the
difference between the total energy of the two shells before and after
the move yields 
\begin{eqnarray}
\label{eq:Ediff}
\Delta E &=& \frac{\G}{12 R^2}\Bigg\{\left(\Delta R^{\prime}-\Delta
  R\right)\Mdm\bigg[2\Mdm +  \\\nonumber
&&\qquad\qquad\qquad  3\left(\Mint+\Delta M_* +\frac{R}{\Delta R^{\prime}}
  \Mext\right)\bigg]-  \\\nonumber 
&& \qquad\qquad 3\Delta R \Delta M_*\left(\Mdm + 2 \Mint+\Delta M_*\right)\Bigg\}\,.
\end{eqnarray}
This can be solved numerically for $\Delta R^{\prime}$, keeping in
mind that $\Mext$ is a 
function of $\Delta R^{\prime}$. In the limit that $\Mext\to 0$, 
equation \ref{eq:Ediff} can be solved analytically for $\Delta R^{\prime}$:
\begin{eqnarray}
\frac{\Delta R^{\prime}}{\Delta R} &=& \frac{1/2 \Mint + 1/3 \Mdm +
  3\Delta M_*
   }{1/2\Mint+1/3\Mdm+3/2\Delta M_*}+\\\nonumber
  &&\quad\frac{(\Delta M_*/\Mdm)(\Mint+1/2\Delta
    M_*)}{1/2\Mint+1/3\Mdm+3/2\Delta M_*} \, .
\end{eqnarray}
In the limit that $\Delta M_* << \Mdm$ the ratio of widths goes to $1$
as expected.  This procedure can be
easily scaled up to a series of interleaved dark matter and stellar
shells.  As the stellar shells move inwards to form a galaxy, they
expand each dark matter shell they cross, thereby slowly moving
dark matter outward.

In the example above, all the energy from the stars is deposited in
the dark matter layer which the stars cross. In our numerical
calculations, the stars deposit their energy in a set of layers
surrounding the layer they cross. The width of this set is
proportional to the current radius and the amount of energy deposited
in each layer is proportional to the mass of that layer. This approximates
the wake created by infalling stellar material, which is responsible
for the forces causing dynamical friction \citep{Weinberg86}. The size
of the wake scales 
as $\G \Delta M_*/\sigma_{\mathrm{dm}}^2 \approx (\Delta M_*/\Mint)
R$. We assume that 
the stars are added by a series of minor mergers, in which each added
stellar shell of mass $\Delta M_*$ constitutes a minor merger. The
mass ratio of such 
mergers is approximately $(\Delta M_*/\Mint)\approx 1/10$, thus setting the
size of the wake.

In a more accurate treatment, there would also be
a diffusive term; each dark matter shell would spread out in radius as
it was on average moved outwards when passed by a stellar
shell.

In the calculations above, we assume no stellar
shells cross each other. This is equivalent to assuming all the
infalling material remains on spherical orbits. This is not true in
reality, especially since galaxies are not spherical but often
triaxial systems, which do not allow purely circular orbits as assumed
above. If
we allowed for triaxial systems in our models and allowed for the
accretion of material on radial orbits, the infalling stars would
deposit energy interior and 
exterior to their final mean orbital radius. This would have an effect
on the final dark matter density profile, but the effect would depend
on the fraction of energy deposited interior and exterior to the final
orbital radius. If all the energy is deposited interior to the final
orbital radius, then more dark matter would be displaced from the
center, leading to a lower central dark matter density. The opposite
is true if the energy is deposited outside the final orbital
radius. 

Additionally, if we relax the requirement that no stellar
shells cross, we must take into account energy deposited in the stars,
not just the dark matter. This will expand the stellar orbits, in the
same way the dark matter orbits are expanded, and lower the stellar
density in the center of the galaxy in the same way the dark matter
density is lowered. Therefore, in order to make a galaxy with a given
stellar density, we would first have to make a more 
concentrated stellar system, and then add stellar clumps which would
undergo dynamical friction against the highly concentrated stellar
system, thereby lowering the central stellar density to the
desired value.  Indeed, observations have
been made of highly concentrated stellar systems at higher redshift
\citep{vanDokkum08, Cappellari09}. In order to make highly
concentrated stellar systems, the galaxies would have to undergo an
early period of dissipational formation. This would also increase the
central dark matter density before the onset of dissipationless
formation, making
the final dark matter density dependent on the relative importance of
dissipational and dissipationless formation mechanisms. Simulations
show that early type 
galaxy formation can be divided into two phases, an initial
dissipational formation of a centrally concentrated system, followed
by accretion via `dry' mergers of additional stellar material
\citep{Naab07, Naab09, Cook09}. If we allowed stellar shell crossing, we would have
to take into account the two phase growth of galaxies and combine the
dissipational and dissipationless models into a single model. This
work attempts to determine the relative importance of the
dissipational and dissipationless formation mechanisms; we are only
concerned with the two extreme formation models, which can be easily
modeled assuming spherical galaxies and infalling material on circular
orbits. The effects of triaxiality and radial orbits would require
combining the dissipational and dissipationless formation mechanisms,
which is left for future work.

Of course, dynamical friction on  incoming stellar clumps is
intrinsically a three-dimensional process 
\citep{TremaineWeinberg84, Aubert06, Aubert07}, and the treatment in
spherical shells 
above is
not intended to accurately mimic the actual assembly of a galaxy via dynamical
friction. Rather, the adopted model is designed to be correct with
respect to the total energy deposited in the dark matter. 
In the numerical calculations presented below, the total binding
energy is conserved. As the width of the stellar shells decreases, the
calculations conserve energy to approximately one part in $10^5$ of
the binding energy 
of the stars in the final galaxy.

%
\subsection{Dissipational Model: Adiabatic Contraction}
\label{ssec:modelAC}
For the dissipational build-up of galaxies we present the following
picture: baryons in the form of gas slowly fall into the centers of
dark matter potential wells, where the gas condenses to form stars.
In this case, the gas radiates away its orbital and thermal energy as it 
falls inwards, so the total energy of the system is not
conserved. However, as long as the gas is
accreted into the center of the galaxy on a timescale that is long compared
to the local dynamical time, the adiabatic invariants of the dark matter orbits
will be conserved. 
In our model for adiabatic contraction, we follow the
prescription of \citet{Blumenthal86} and again assume circular orbits
for the dark matter and a spherically symmetric mass
distribution. Instead of energy being conserved, the
adiabatic invariants of the dark matter orbits are conserved.  For
periodic orbits, $\oint\,p\mathrm{d}q$ is an adiabatic invariant,
where $q$ is a coordinate and $p$ is its conjugate momentum.  For a
particle in a circular orbit at a radius $r$ around a spherical mass
distribution $M(r)$, we take the conjugate momentum to be the angular
momentum and its corresponding coordinate, the angular position.  The
adiabatic invariant is then:
\begin{equation}
  \label{eq:adiabatInvariant}
  J^2 = \Big(\oint\,\sqrt{M(r)r}\mathrm{d}\theta\Big) \propto rM(r)\,.
\end{equation}
Therefore, if the
mass interior to the orbit, $M(r)$, increases, the orbital radius must
decrease. Following the same set-up as in the previous model, we start
with a dark matter shell of constant density and mass $\Mdm$ directly
interior to an infinitesimally thin shell of stars/gas of mass
$\Delta M_*$. After the baryons move interior to the dark matter layer, 
the inner radius of the dark matter shell becomes $R^{\prime}= R \, \Mint / (
\Mint + \Delta M_* )$ and the width of the layer becomes 
\begin{equation}  
  \label{eq:adContractWidth}
  \Delta R^{\prime} = \frac{(\Mint+\Mdm)(R+\Delta R)}
  {\Mint+\Mdm+\Delta M_*}
  - R^{\prime}\, .
\end{equation}
Overall, the dark matter shell moves inwards and becomes thicker or
thinner depending on the mass ratios of the dark matter shell, the
baryon shell, and the mass interior to both. As in the previous model,
two shells only interact when their orbits cross, so this prescription
for interchanging two 
shells can be scaled up to many shells.

If the constraint of circular orbits is removed, the adiabatic
invariant is no longer $r M(r)$. Using N-body simulations,
\citet{Gnedin04} show that $M(\bar{r}) r$, where $\bar{r}$ is the
orbit-averaged position, is a good proxy for the adiabatic
invariant. If this quantity is conserved for isotropic orbits, the
prescription for adiabatic contraction remains the same.  However, the
prescription above will overestimate the amount of adiabatic
contraction if the orbits are radially biased \citep{Gnedin04}.

\subsection{Models Fit to Example of Massive Galaxy}
\label{ssec:example}
The difference in the dark matter density in the central regions of 
a galaxy for each model is clearly shown by the comparison of the velocity
profiles of model galaxies. In the top panel of Figure \ref{fig:vcirc},
the circular 
velocity curves for both models are compared to the velocity curve
produced by the stars alone. The models shown in Figure
\ref{fig:vcirc} are taken to fit very massive galaxies and we have
adjusted \MsL{} such that the central velocity dispersions of both
models are the same. The central velocity dispersions are computed
assuming the stars are on isotropic orbits. In both cases, the stars
dominate in the
very central regions, but the differences in dark matter significantly
affect both curves.  The lower panel of Figure \ref{fig:vcirc} shows the ratio
of the dark matter to stellar matter projected densities as a function
of radius for both models. In reality, since galaxies certainly form by a
combination of dissipational and dissipationless methods, the velocity
curves and the dark matter to stellar matter density ratios will fall
somewhere in between the two extreme toy models examined here. 

\middfig{\figvcirc}

In order to compute the two models, we require a set of input
parameters describing the initial matter distribution and the final
stellar distribution.  If we assume the initial distribution of both
components fit a single NFW profile, then the two required input
parameters are the total mass of the halo and the initial
concentration. Assuming that the galaxy is 
spherically symmetric, the final stellar mass
distribution is completely described by the surface brightness profile
of the galaxy
and a stellar mass-light ratio. The luminosity profile can be
directly observed and \MsL{} can be determined from observations of the
central velocity dispersions of
galaxies. Since stars dominate the mass in the central regions, they
also are the dominant contribution to the central velocity
dispersion. Thus, the
observational input parameters for these models are the total
luminosity of the galaxy, the luminosity profile (in the case of
Sersic profiles, the necessary terms are the Sersic index, $n$, and the
half-light (effective) radius, $\Reff$),  and the central velocity
dispersion. The free 
parameters are the total halo mass, the initial NFW concentration, and
the stellar mass-to-light ratio (which we assume to be constant
throughout a given system).  Therefore, for a given galaxy, the
two models will by construction have the same luminosity and central
velocity dispersion, but 
the total halo masses and the mass-to-light ratios will be different,
and in some cases, outside the bounds set by other
observational and 
model constraints. Table \ref{table:models} gives the parameters for
the dissipational and dissipationless models which best fit the strong
lensing cluster MS-2137-23 discussed in \S\ref{ssec:strongLens}.

\subsection{Minimum Radius for Galaxies Formed by Dissipationless
  Mergers}
\label{ssec:minRad}
In order for a galaxy to form by purely dissipationless processes, the
incoming stars must deposit their orbital energy in the dark matter
halo. Therefore, the dark matter halo must initially have sufficient
\emph{binding} energy to give to the infalling stars. The stars
in the final galaxy can have no more binding energy than the initial
dark matter halo. This sets a lower limit on the size of a galaxy
formed by purely dissipationless processes. A galaxy formed by
dissipationless accretion would approach the minimum size and would
have very little dark matter in the center, which is in agreement with
observations mentioned in \S\ref{sec:intro}. However, if 
dissipational processes play a dominant role in
galaxy formation, there is no minimum size for galaxies, as the
infalling baryons can dissipate all of their orbital and thermal energy.

Given a cluster mass, concentration, and final galaxy stellar mass,
and Sersic index, 
a lower limit on the galaxy's effective radius, $\Reff$, can be obtained by
setting the change in binding energy of the dark matter halo equal to change
in binding energy of the stars and baryons in the cluster. For
galaxies with $L\approx 10^{11}\ \Lsun$ in the $K$-band, the minimum
effective radius for a dissipationlessly formed galaxy yields a
reasonable minimum for the effective radius of observed early type
galaxies. \citep{Bernardi03I}.

We should note that there is another limiting radius for the formation
of accreted halos. A satellite may be destroyed by
tidal shocks before it reaches the energetically allowed minimum
radius, thereby depositing stars in the outer region of
the growing galaxy \citep{Kormendy77, Gnedin99, Wetzel09}. This
process is difficult to compute but gives a limiting 
radius comparable to the dynamical friction limits computed above. For
the most massive systems (BCGs), the tidal shock limiting radius is
more severe than the dynamical friction one, so we can expect that the
dark matter to stellar matter ratio in such systems will be higher
than in more moderate mass galaxies. 

\section{Comparisons to Observations of BCGs}
\label{sec:ComparetoObs}

One test of the dissipational and dissipationless models of galaxy
formation is the build-up of
brightest cluster galaxies (BCGs). BCGs offer a good comparison sample
for these 
spherically symmetric test models for several
reasons.  First, BCGs almost always sit at or nearly at the center of
their host cluster. Therefore, they are also centered in a dark matter
halo, so there are no contributions to the potential from an
off-center halo, not included in our models. Furthermore,
BCGs represent a uniform sample, so much so that they have been suggested
as standard candles \citep{Postman95}, allowing comparisons to be made
to the entire population instead of individual galaxies. Finally,
BCGs are thought to have been formed by a series of galaxy mergers
during the build-up of clusters \citep{OstrikerHausman77, Nipoti04,
  Cooray05}, making them 
good candidate systems in which to observe dissipationless galaxy formation. 

\subsection{Scaling Relations for Input Parameters}
\label{ssec:scaling}

In order to compare the models to observations, we normalize the
models to $K$-band and $r$-band data of BCGs. Given the luminosity of a
BCG and choosing a constant stellar mass-to-light ratio for the models,
we use empirical relations to derive the cluster and the BCG
properties. From \citet{LinMohr04}, the cluster mass is related to the
observed 
$K$-band BCG luminosity by
\begin{equation}
  \label{eq:MclLumLin}
  \frac{\Lbcg}{10^{11}\h^{-2}\ \Lsun} = \left(4.9 \pm
  0.2\right) \left(\frac{\Mcl}{10^{14}\ \Msun}\right)^{0.26\pm0.04} \, .
\end{equation}
The luminosities of BCGs are 6-10 times brighter than $\mathrm{L}_*$
in the $K$-band. For galaxies typically in clusters, $M_{K*} =
-24.34$ \citep{LinMohr04}, including the BCG, or about
$1.16\times10^{11}\ \Lsun$ in 
the $K$-band. 
From the cluster mass, the cluster virial radius, $r_{200}$, can be
calculated assuming the critical density  $\rho_{\mathrm{cr}} = 1.36\ \h^2
\times10^{11}\ \Msun\ \mathrm{Mpc}^{-3}$. Simulations have shown that
the 
dark matter halo concentration,  $c = r_{\mathrm{200}}/r_{\mathrm{NFW}}$, scales
approximately with the
mass as \citep{Neto07}
\begin{equation}
  \label{eq:concentration}
c = 4.67 \left(\frac{\Mcl}{10^{14}\ \Msun}\right)^{-0.11}\, .
\end{equation}

Equations \ref{eq:MclLumLin} and \ref{eq:concentration} set the
initial conditions for the models. 
The properties of the stellar component of the BCG can also be derived
from \Lbcg.  We assume that the BCGs are well-modeled by a single
Sersic profile \citep[$\mathrm{I}(R)\sim
 \exp(R^{1/n}),$][]{Sersic69}, ignoring the ICL
and outer components of the BCG \citep[see][]{Gonzalez05}. The
two-dimensional surface
brightness profile defined by Sersic can be deprojected numerically into a
three-dimensional luminosity density profile, assuming the galaxy is
spherically symmetric.  This numerical deprojection is well-approximated by the
analytic formula \citep{LimaNeto99}:
\begin{eqnarray}
 \label{eq:Sersicdensity}
 \rho_*(r) &\propto& (r/\Reff)^ {1-1.188/(2n)+0.22/(4n^2)} \nonumber\\
 &&\exp\left((0.327-2n)(r/\Reff)^{1/n}\right)\,,
\end{eqnarray}
where $n$ is the Sersic index ($n=4$ for a de Vaucouleurs profile),
and \Reff{} is the half-light radius of the
surface brightness profile. Observations show that the Sersic 
properties of BCGs are correlated with the galaxy's luminosity. Using data from
\citet{LinMohr04} and \citet{Graham96}, the luminosity can be related
to the half-light radius and the Sersic index by
\begin{eqnarray}
  \label{eq:LReffn}
  \log\Reff &=& -10.30 +
  1.01\log\left(\frac{\Lbcg}{\Lsun}\right) \quad \mathrm{and}\\
  n &=& 2.9\log\Reff+1.98\, ,
\end{eqnarray}
which are in good agreement with scaling relations from
\citet{ValeOstriker08}, \citet{Bernardi07} and \citet{Desroches07}.

Also modeled in each galaxy is a central supermassive black hole,
which adds a minor correction to the velocity dispersion of the
galaxy.  The black hole mass is determined from the galaxy luminosity
by the relation \citep{Graham07}: 
\begin{eqnarray}
  \label{MbhLumGraham07}
  &&\log\left(\frac{\Mbh}{\Msun}\right) = \\\nonumber 
  &&\quad (0.95\pm 0.15) \log\left( \frac{\Lbcg}{10^{10.91}\
      \Lsun{}_{\mathrm{,}K}} \right) +  (8.26\pm0.11) \,.
\end{eqnarray}
Thus, by
supplying a galaxy luminosity and a stellar mass-to-light ratio,
we can obtain all the other input parameters needed to compare the
dissipational and dissipationless models to observations of BCGs.


\subsection{The $L$--$\sigma$ Relation}
\label{ssec: FaberJacksonBCGs }

The innermost probe of the mass profile is the central velocity
dispersion of a galaxy.  Elliptical galaxies fall on the
fundamental plane \citep{DjorgovskiDavis87, Dressler87} and one
projection of the plane is the Faber-Jackson 
relation: the
relation between a galaxy's luminosity, $L$, and velocity
dispersion, $\sigma$ \citep{FaberJackson76}. In the $K$-band, the
Faber-Jackson relation observed by \citet{Pahre98} is 
\begin{equation}
\label{eq:FJ}
M_K = -10.35\pm0.55\log\sigma_0\, .
\end{equation} 
For BCGs, \citet{Lauer07} show that the
velocity dispersion saturates at 
high luminosities, leading to the relation
\begin{equation}
\label{eq:lauerFJ}
M_V = -2.5 (6.5\pm1.3) \log\left(\frac{\sigma}{250\ \kmps}\right) -
22.45 \pm 0.18 \,. 
\end{equation} 
\citet{Desroches07} find a similar relation. In order to
compare to the $L$--$\sigma$ relation for BCGs, we normalize our
models to observed BCGs using the scaling relations described in
\S\ref{ssec:scaling}. The line-of-sight central velocity dispersions
averaged over an aperture of $1.64\ \mathrm{kpc}$ are then calculated
for both the dissipationless and dissipational models.  The comparison
to the $L$--$\sigma$ relation from \citet{Lauer07} is shown
in Figure \ref{fig:Lsigma}. Although both the dissipational and
dissipationless models are slightly steeper than the
$L$--$\sigma$ relation found by Lauer, the dissipationless
model has a slope that more closely matches the observed
$L$--$\sigma$ for BCGs. To match each model
to the observations, the stellar mass-to-light ratios can be
adjusted. For the dissipational models, the best fit \MsL{} in the
$K$-band is 1.43, while for the dissipationless models the best fit to the 
 $L-\sigma$ relation is for $\MsL{}=2.40$.  These are equivalent to
 stellar mass-to-light ratios of 7.57 and 12.71 in the $V$-band, assuming
 $V-K = 3.31$ for the BCG population. In the $K$-band,
 the stellar mass to 
 light ratio measured by \citet{Cole01} is 
$0.73$ for a Kennicut IMF and $1.32$ for a Salpeter IMF.
 
The stellar mass-to-light ratios derived for these models are simply
the mass in stars needed to reproduce the dynamics (in this case, the
central velocity dispersion) divided by the total observed luminosity
of the galaxy. Although the dissipationally formed galaxies were
brighter at high redshift due to star formation, we are only concerned
with the $z\approx0$ luminosity and dynamical state of the
galaxy. This corresponds to the luminosity of the evolved population;
therefore, we have implicitly included passive evolution in the
dissipational model and do not need to passively evolve the \MsL{}
values derived above.

However, the mass-to-light ratios are sensitive to the empirical
scaling relations.  For example, if the relation for $\Reff(\Lbcg)$ is
replaced with that derived by \citet{Bernardi07} for galaxies fit by
de Vaucouleurs profiles ($n=4$), the best fit stellar mass-to-light
ratios become $0.98$ and $1.76$ for the dissipational and
dissipationless models respectively.

Additionally, the above calculations rely on the scaling relation
between the total halo mass and the BCG luminosity. Instead, we can
assume that the cluster is built
hierarchically out of galaxies formed at $z\approx2$. At $z=2$, the
concentration of a dark matter halo is a weak function of halo mass;
\citet{GaoNavarro08} find that $c \propto \Mhalo^{-0.031}$. The
fraction of mass in a halo which will form stars is given by
$M_*/\Mhalo \propto \Mhalo^{-0.26}$ \citep{Lin03,Bode09}. Using these
relations for the concentration and stellar mass, the best-fit \MsL{}
values become $0.69$ and 
$1.55$ for the dissipational and dissipationless models respectively.
These values are in better agreement with the measured values given above. 

It is not surprising that neither value for \MsL{} 
 can be discarded based on the Faber-Jackson 
 relation. In the central regions of the galaxies, both models are
 stellar-dominated, as shown in the lower panel of Figure
 \ref{fig:vcirc}.  Therefore, even though the dark matter density 
 can differ by more than a factor of 10, it only makes up $\sim10\%$ of
 the total mass in the central regions, and thus does not determine the
 central dynamics.

\middfig{\figLsigma}

\subsection{Microlensing Optical Depth}
\label{ssec:microlens}

Microlensing of quasars, which has been observed in
multiply imaged systems \citep[and references
therein]{Wozniak00, Wambsganss06},
in principle provides a probe of the mass function
of microlenses (MACHOS, stars, or dark matter substructure), and the
density of these microlenses relative to a 
smooth background density \citep{SchechterWambsganss02,Dobler07,Pooley09}. Searches in the Sloan 
Digital Sky Survey have found $\sim220$ 
strongly lensed quasars \citep{Inada08}, which are lensed by
individual galaxies or entire clusters. In the following, we use the
best-fit models for BCGs as an
example to show the expected differences in microlensing results
between the dissipational and dissipationless formation models.

The difference in stellar mass-to-light ratios between the dissipational and dissipationless models leads to 
differences in the microlensing optical depth.  The optical depth,
$\tau$, is proportional to the number density of lenses, stars in
this case, times the Einstein radius, $\theta_{\mathrm{E}}$, of
each lens. Assuming that the distance between lens and source is
 large compared to the size of the
 galaxy, the microlensing optical
 depth is \citep{Paczynski86}: 
\begin{equation}
  \label{optDepthML}
  \tau = \Sigma_* \frac{4\pi \mathrm{G}}{c^2}
  \frac{D_{\mathrm{ls}}D_{\mathrm{l}}}{D_{\mathrm{s}}}\,, 
\end{equation}
 where $\Sigma_*$ is the projected stellar density and $D_i$ are the
 angular diameter distances to the lens, to the source, and between the lens
 and the source. If $(D_{\mathrm{ls}}D_{\mathrm{l}})/D_{\mathrm{s}}= D$
 and the Sersic index of the lens galaxy is 
 assumed to be 4.0, then the microlensing optical depth at the
 half-light radius is
 \begin{eqnarray}
  \label{tau1gpc}
  \tau &=& 1.33\times 10^{-2} \left(\MsL\right)_K \left(
    \frac{\Lbcg}{10^{11} \ \Lsun}\right) \\\nonumber
  &&\qquad\qquad \left( \frac{\Reff}{10 \ \mathrm{kpc}} \right)^{-2} \left(
    \frac{D}{0.5 \ \mathrm{Gpc}}\right) \,.
 \end{eqnarray}
For the dissipational and dissipationless models that best fit BCGs,
the ratio of the microlensing optical depths equals the ratio of
\MsL{}, or $1.43/2.40 = 0.60$.  In most cases, the microlensing
optical depth at the position of the 
image is of order unity. Occasionally, individual microlensing events can be
observed. 

Additionally, the relative
density of smoothly distributed matter (dark matter) to microlenses
(stars) can be probed
\citep{SchechterWambsganss02, Dobler07, Pooley09}. Using
the dissipational and dissipationless models shown in Figure
\ref{fig:vcirc} as an example
($L=6.0\ L_*$), the fraction of dark
matter to total matter along 
a line-of-sight is 0.81 at 0.1 \Reff{} and 0.98 at 1.0 \Reff{} for the
dissipational model. For the
dissipationless model, the ratios at 0.1 \Reff{} and 1.0 \Reff{} are 0.48
and 0.97, respectively. These large differences should be measurable in
microlensing studies of multiply-imaged quasars.

\subsection{Strong Lensing}
\label{ssec:strongLens}

\middfig{\figlenses} 

Strong lensing measurements provide a clean method of probing the
total mass
distribution of BCGs and their host clusters. \citet{SandTreu04}
present observations of radial and tangential arcs for six
clusters acting as lenses. The positions of the radial and tangential
arcs are given by the solutions to
\begin{eqnarray}
  \label{eq:lenseqn}
  0 &=& 1 -
  \frac{\mathrm{d}}{\mathrm{d}R}\frac{M_{\mathrm{proj}}(R_{\mathrm{rad}})}{\pi
    R_{\mathrm{rad}}} 
  \quad\mathrm{and}\\\nonumber 
  0 &=& 1 - \frac{M_{\mathrm{proj}}(R_{\mathrm{tan}})}{\pi R_{\mathrm{tan}}^2} \,, 
\end{eqnarray}
where $M_{\mathrm{proj}}(R)$ is the projected mass interior to $R$ scaled by the
critical surface density, 
\begin{equation}
  \label{eq:sigmaCR}
  \Sigma_{\mathrm{cr}} = \frac{c^2}{4\pi
    G}\frac{D_{\mathrm{s}}}{D_{\mathrm{ls}}D_{\mathrm{l}}}\, . 
\end{equation}

Together, the radial and tangential lenses constrain the slope of
the density profile and its normalization. \citet{SandTreu04} use the
lensing information as well as the velocity dispersion profile of the BCG to
create density models for the stars and the dark matter in each
lens. They find that the mean inner dark matter density profile
for six lensing clusters is
$r^{-0.52\pm0.3}$, significantly shallower than the NFW profile. 

We repeat the analysis of \citet{SandTreu04},
fitting our dissipational and dissipationless models for the 
dark matter profile to the three clusters with both radial and
tangential arcs. As in the previous section, we assume that each BCG
in the center of the cluster is built hierarchically, either a series of purely
dissipationless mergers of smaller stellar systems or by the
dissipational accretion of gaseous streams which lead to in situ star
formation.  The dissipationless model for formation will yield a lower dark
matter density in the center of the cluster, while the dissipational
model for formation will lead to adiabatic contraction of the dark
matter. At $z\sim2$, 
both dissipational and dissipationless formation mechanisms will be
important, but the ratio between the two mechanisms is unknown, and by
comparing data to the purely dissipationally and dissipationlessly
formed BCGs, we hope to constrain how much each mechanism contributes
to galaxy formation.

For the dissipational and dissipationless models, the fixed input
parameters for the models are the BCG 
luminosity, half-light radius, and Sersic index ($n=4$). The free
parameters are the total cluster mass, \Mcl,  the dark halo concentration,
$c$,  and \MsL{}. By randomly selecting these input parameters from a
reasonable range, we can find both dissipational and
dissipationless models that are within $1$
and $2\sigma$ of the measured central
velocity dispersion and radial and tangential arc
locations. Projections of these points in \Mcl--$c$
space and \Mcl--\MsL{} space are shown in Figure \ref{fig:lenses}.  The
cluster Abell 383 does not have any models which lie within $1\sigma$
of the observations, so the $2\sigma$ models are plotted instead. These
projections show that the best-fit models lie on a tight
relation between \Mcl{} and $c$. From equation \ref{eq:concentration},
halos in this mass range should have a concentration between 3.2 and
5.0, eliminating most of the best-fit models for MS 2137-23 and
RXJ-1133. In the case of MS 2137-23, this eliminates
almost all of the dissipationless models (squares). However, weak
lensing measurements of MS-2137-23 predict a concentration about twice
as large as simulations, thereby only eliminating a few models
\citep{Gavazzi03}. 

Also illustrated in
Figure \ref{fig:lenses} is the difference in \MsL{} between the two
models.  The heavy lines indicate the median and
$25-75^{\mathrm{th}}$ percentile (SIQR) for the best-fits for each toy
model. As with the comparison to the $L$--$\sigma$ relation, the
dissipational models  
have lower \MsL{} values than the dissipationless models.  For example,
the median $(\MsL)_V$ ratios for MS 2137-23 are 
$4.3\pm2.4$ and $1.2\pm0.6$ for the dissipationless and dissipational
models, respectively. For RX-J1133, $(\MsL)_B=2.1\pm0.8$ for the
dissipational models and $(\MsL)_B=5.2\pm1.7$ for the dissipationless
models. The best-fit dissipational models
all have an \MsL{} below 2.5(3.5) in the $V(B)$-band. Assuming passive
evolution, the expected value 
of \MsL{} is $(\MsL)_B\approx4.1\pm0.95$ at
$z\sim0.35$ \citep{TreuKoopmans04,vanderWel04, Treu06}. The purely
dissipational model for RX-J1133 is therefore marginally
inconsistent with expected \MsL{} values. The shaded regions in Figure
\ref{fig:lenses} show the expected ranges for the stellar
mass-to-light ratios for the three clusters; MS 2137-23 is consistent
with both the dissipational and dissipationless models. Neither the dissipational nor the dissipationless models represent an adequate fit to Abell 383; however, the \MsL{} values for the dissipational models within $2\sigma$ of the observations are in better agreement with the expected \MsL{} value.

Taking MS 2137-23 as a specific example, Figure \ref{fig:msbestfit} plots
the stellar and dark matter density of the two models. Both models
selected have tangential and radial arcs and velocity dispersions
within the error bars of the observations. The best-fit model
parameters and the observed model parameters are given in Table
\ref{table:models}. For the dissipational model, the
dark matter density dominates over the stellar density at all radii.
From \citet{SandTreu04}, the best-fit inner slope of the dark matter
profile is 0.57.  This is shown along with the $2\sigma$
error bars. The inner slope derived by \citet{SandTreu04} depends on
the concentration remaining fixed at 400 kpc. If the concentration is 
allowed to vary, the best-fit inner slope will generally increase by
0.15, bringing it closer to the dissipationless model. However, the
dissipationless model also depends strongly on concentration and there
exist choices for \Mcl, \MsL{}, and $c$, which fit the observations
equally well, such that the dissipationless model almost matches an
NFW profile.  

\middfig{\tablemodels}
\middfig{\figmsbestfit}

As illustrated by the strong lensing, the differences between the two
models for BCGs are small.  This is
due to the fact that the ratio between the stellar mass of the galaxy
and the dark matter halo mass is very small, on the order of
$0.002$. Figure \ref{fig:Mratio} shows the ratio of the dark matter
to stellar mass inside $0.25\Reff$  as a function of galaxy luminosity
using the scaling relations from \S\ref{ssec:scaling}. As the
luminosity and cluster mass increase, the differences
between the two models and the initial NFW profile become
smaller. Therefore, the differences between the dissipational and
dissipationless methods of
galaxy formation will be most pronounced in smaller dark matter halos,
such as fossil groups and isolated ellipticals.  In these cases the stellar
component of the central galaxy is much larger relative to the halo
component, making the differences between the dissipational and
dissipationless models more pronounced.

\middfig{\figMratio}

\section{Comparison to SAURON Data}
\label{sec:sauron}

Figure \ref{fig:Mratio} illustrates that the total mass-to-light
ratio is an increasing function of galaxy luminosity. This is in
agreement with the trend found by the SAURON project
\citep{Cappellari06}, which uses integrated-field spectroscopic
observations of 25 E/S0 galaxies. Using these observations and stellar
population models to determine the stellar mass-to-light 
ratios for their sample of galaxies, \citet{Cappellari06} find that the total(dynamical) \ML{} is
consistently larger than \MsL{} and that this difference increases with
increasing stellar mass. This trend is shown in Figure
\ref{fig:sauron}. The `$\times$'-symbols  denote the SAURON data and the
shaded region is the best fit. The stellar mass-to-light ratio
($I$-band) is never larger than 3.4 for the brightest galaxies. The
best fitting line is given by \citep{Cappellari06}
\begin{equation}
  \label{eq:SAURON}
  \left(\ML\right)_I = (2.35\pm0.19)\left(\frac{L_I}{10^{10}L_\odot}\right)^{0.32\pm0.06}\,.
\end{equation}
This fit ignores the galaxy (M32) at $M_I \sim -17.5$.

\middfig{\figSAURON}

We can compare our dissipational and dissipationless models to the
SAURON data to determine whether the models recover the trend given by
equation \ref{eq:SAURON}. As in the previous section, we assume that
each of the 25 galaxies in the \citet{Cappellari06} study is formed
either by purely dissipational or purely dissipationless processes. We
fix the stellar mass-to-light ratio, the total $I$-band luminosity,
and the
effective radius (\Reff{}) for
each galaxy to the values given in \citet{Cappellari06}. We then vary
the dark matter halo mass for the dissipational and
dissipationless models of each galaxy until the velocity dispersion
within \Reff{} for both models matches the value reported in
\citet{Cappellari06}. Since dissipational formation increases the dark matter content in the center of a galaxy
compared to the dissipationless model, a smaller total halo mass is required to
recover the same central velocity dispersion in the dissipationally
formed galaxies than in the dissipationlessly formed galaxies. The
differences in halo mass lead to differences in dynamical \ML{}, which
are shown in Figure \ref{fig:sauron}. Both the dissipational and
dissipationless models reproduce the same trend in \ML{} with luminosity
as is shown in the SAURON data. However, the dissipationless models
yield slightly lower \ML{} values than the dissipational models, leading
to better agreement with the observed values. The standard deviation
of the SAURON points around the best fit line is 0.11. The
standard deviation of the dissipationless model points (squares)
around the best-fit line to the SAURON data (dotted line) is
0.14. The same value for the dissipational models is 0.25. However,
both the dissipational and dissipationless models have \ML{} values
higher than those from the SAURON data. Therefore, no combination of
these models will yield the measured SAURON galaxies. However, both
the dissipational and dissipationless models used here assume the
stellar orbits are 
isotropic and that the galaxies are spherically symmetric. Both of
these assumptions will affect the model-calculated \ML{} values; in
the case of rotating galaxies, the calculated \ML{} values will be
lowered and possibly brought into better agreement with the SAURON
observations. 

\section{Example Galaxy: NGC 4494}
\label{sec:ngc4494}

NGC 4494 is an ordinary elliptical galaxy with a $B$-band luminosity of
$2.37\times10^{10}\ \Lsun$. Because it is an isolated
galaxy instead of a BCG, the mass ratio between the dark matter halo
and the stars is smaller and, therefore, the difference between the
dissipationless and dissipational models of formation will be larger
than those found for BCGs in massive clusters. As with the BCGs, the
input model parameters for NGC 4494
can be constrained by observations.  

\subsection{Velocity Dispersion using Planetary Nebulae}
\label{ssec:PNe}
Planetary nebulae have been established as a good mass tracer in the
outer regions of galaxies.  These observations provide a good
comparison case for our extreme models of galaxy
formation. \citet{Napolitano09} measure positions and velocities of 
planetary nebulae out to $\sim7\ \Reff$ in the elliptical NGC
4494, probing the velocity dispersion for the galaxy much farther out
than the central velocity dispersion. At these 
large radii, the dark matter will be comparable to, and dominate over,
the stellar matter (see Figure \ref{fig:PNeDMdensity}), thus
emphasizing the differences between the dissipational and
dissipationless models. By fixing
the luminosity ($L_B = 2.37\times10^{10}\ \Lsun$), the effective
radius ($\Reff=3.68\ \mathrm{kpc}$),
and the Sersic index ($n=3.30$),  of
the model galaxies to observations
from \citet{Napolitano09}, we can fit our
dissipational and dissipationless models to the 
planetary nebulae velocity dispersion curves by varying the total halo
mass
and the stellar mass-to-light ratios. These fits also include a point
for the central
velocity dispersion, $\sigma=150.2\pm3.7\ \kmps$, as reported in the
Hyperleda\footnotemark\footnotetext{http://leda.univ-lyon1.fr} 
database \citep{Paturel03}. The results are shown in Figure
\ref{fig:PNe}. For the dissipational
model, $(\MsL)_B=2.97$, while the
dissipationless model has a mass-to-light ratio of $3.96$.  The total
halo masses are $6.0\times10^{11}\ \Msun$ for the dissipational model
and $1.0\times10^{13}\ \Msun$ for the dissipationless model. The
slightly 
poorer quality fit for the dissipationless model is due to the fact
that the galaxy's effective radius is close to the minimum allowed for
the galaxy to form via dissipationless mergers ($\sim 
2.7\ \mathrm{kpc}$), as discussed above in \S\ref{ssec:minRad}. As the
galaxy approaches this minimum size, there is insufficient binding energy
in some of the central dark matter layers to allow the stellar layers
to cross. The fit of the
dissipationless model could be improved if we relaxed our
model requirement that all the energy from the stars is deposited
locally, instead allowing more energy to be deposited in the outer
regions of the halo. This could be the case if the orbits of the
in-falling material were radial orbits instead of perfectly
circular orbits as assumed in this work.

Although the dissipational model shown in Figure \ref{fig:PNe} provides a better fit to the data, the $B$-band mass-to-light ratio required for the dissipational model is significantly lower than the value derived from stellar
population models, $4.3\pm0.7$ \citep{Napolitano09}. Thus, a purely
dissipational formation of NGC 4494 appears to be ruled out at the
$\sim1.9\sigma$ level. 

\middfig{\figPNe}

Although adjusting the stellar mass-to-light ratio eliminates the
differences in the velocity dispersion profile for these two models,
the difference in dark matter
density between them remains large (see Figure 
\ref{fig:PNeDMdensity}). The inner dark matter density profile 
for the dissipationless model follows $\sim r^{-0.2}$, while that of the
dissipational model follows $\sim r^{-1.7}$, making the central dark
matter densities in the two models very different. The slope
index of the dark matter in the dissipational model is in good
agreement with that predicted for a final isothermal mass distribution
in \S\ref{sec:intro}. 

\subsection{Dark Matter Annihilation}
\label{ssec:darkAnnihilation}

Although only available in the Milky Way, one direct method of
probing WIMP dark matter currently being explored
is the observation of gamma rays from the self-annihilation of WIMP dark
matter particles \citep{Stoehr03,Colafranceso06,Diemand07}. The
signal strength from such annihilations will be
proportional to $\rho_{\mathrm{dark}}^2$.  Assuming a smooth distribution of
dark matter, the ratio of the annihilation signal strength within an
aperture of $\Reff$ for the dissipational and dissipationless models of
NGC 4494 is around $1890$, similar to what would be expected for a
Milky Way sized halo. Although not observable today, the Fermi
gamma-ray space 
telescope hopes to measure the dark matter annihilation signal from
our own galaxy. The large difference in signal strength between the
dissipational and the dissipationless models calculated here dominates
over the boost in signal strength expected from unresolved
substructure, which is of order $10$ \citep{Strigari07,Kuhlen08},
providing another possible test of the formation history of the
stellar component of galaxies. 

\middfig{\figPNeDMdensity}

\section{Conclusions}
\label{sec:conclusions}

We have shown that the two extreme cases for the assembly of the
stellar content of galaxies lead to large differences in the dark
matter density profiles of galaxies, assuming that the initial halo 
conditions are well-described by N-body simulations. The stellar mass
density dominates over the dark matter 
density in the central regions of both models; the dark matter
density in the dissipational models can be as much as two orders of
magnitude lower at $r\approx1\,\mathrm{kpc}$ than the dark matter
density in the dissipational models. However, because
galaxies are undoubtedly built up by both dissipational and
dissipationless accretion, most observations will not easily distinguish 
between these two models.  For example, although the best-fit models for
BCGs have different stellar mass-to-light ratios, neither is outside the
acceptable range of values from stellar population models and
observations.  Strong gravitational lensing observations of
BCGs and their host clusters show that the dissipational formation
models have \MsL{} values that are marginally too low compared to
those expected for passively evolving ellipticals. For RX-J1133, the
median $(\MsL)_B = 2.1\pm0.8$ and $5.2\pm1.7$ for the
dissipational and 
dissipationless models, respectively, while the expected
$(\MsL)_B$ from 
passive evolution is $4.1\pm1.0$ \citep{TreuKoopmans04}. This
discrepancy between \MsL{} values 
marginally rules out a purely dissipational formation history for
BCGs, in agreement with both theoretical expectations and
other observational evidence. Observations of strong
lensing by BCGs have been used to 
effectively rule out dark matter density profiles as steep and steeper
than an NFW \citep{SandTreu04}, further strengthening arguments
against 
a purely dissipational formation for the stellar component of BCGs.
Although extreme values for the 
concentration ($\sim10$) 
and $(\MsL)_B$ ($\sim1.0$) are allowed by the lensing and dynamics data, the
dissipational model can be ruled out for a more constrained and
plausible set of model parameters.

Both models adequately reproduce the trend of increasing total \ML{}
with galaxy luminosity for E/S0 galaxies, observed using
integrated-field spectroscopy by the SAURON project
\citep{Cappellari06}. However, 
the lower \ML{} values found for the dissipationless models are in
better agreement with the data.

Constraints on the stellar mass-to-light ratios can also be used to
exclude the purely dissipational model of galaxy formation in the case of
the isolated elliptical, NGC 4494. Fitting the dissipational and
dissipationless models to observations of planetary nebulae yields
$(\MsL)_B$ values of $2.97$ and $3.96$ for the dissipational and
dissipationless models, respectively. Compared to $4.3\pm0.7$, the
$(\MsL)_B$  
inferred from stellar synthesis models, the purely
dissipational model can be ruled out at the $1.9\sigma$ level. 

Since the change in the dark matter density for both models is
directly related to the change in the central mass of the halo, the
larger the stellar component is relative to the dark matter halo, the
larger the
differences between the dissipational and dissipationless extremes
will be. Therefore, instead of examining the properties of BCGs, we
propose looking for the differences between dissipational and
dissipationless formation mechanisms using the brightest galaxies of
fossil groups 
and isolated elliptical galaxies. The large differences attainable in
this mass range of galaxies is clearly illustrated by the study of NGC
4494. The dark matter density profiles 
shown in Figure \ref{fig:PNeDMdensity} have inner slope indices of
$\alpha\approx0.2$ and $1.7$ for the dissipationless and dissipational
models, respectively. The differences in the dark matter density
profiles for galaxies in this mass range are significant enough
that they could be probed by galaxy-galaxy weak
lensing studies, provided the difference in dark matter slopes is not
removed by averaging over many galaxies with different formation
histories. Finally, the difference in dark matter content between the
dissipational and dissipationless models yields differences in the 
signal strength from dark matter annihilation of order $\sim1890$, far
larger than the boost factor expected from the unresolved dark matter
substructure in the Milky Way halo.

The focus of this paper has been the energetics of the
dissipational and dissipationless galaxy formation mechanisms, not
the mechanisms themselves. For dissipational galaxy formation, we have
assumed that baryons cool and condense in the center of halos, leading
to adiabatic contraction of the surrounding dark matter. This behavior
has been confirmed in cosmological simulations. Although simplified,
the model presented here is correct, on average, for more complicated
galaxy formation scenarios, including major as well as minor mergers,
and accretion from filaments instead of spherical shells \citep{Gnedin04}.

The physical mechanism we propose for dissipationless galaxy formation is the
dynamical friction of small stellar clumps against a smooth dark
matter background. In the models used here, we assume circular orbits
for the incoming stellar material. The inclusion of radial orbits and
non-spherical galaxies is left for future work, as it requires
modeling a combination of dissipational and dissipationless
formation mechanisms. We assume that the build-up of
the galaxy occurs via a series of small, minor mergers
\citep{Bezanson09, Cook09, Naab09}, not allowing
for equal-mass mergers, which more violently disrupt the
system. Indeed, it has been shown in dissipationless N-body
simulations that equal-mass merger remnants will retain the profile of
the steepest progenitor \citep{BoylanKolchinMa04, Dehnen05, Kazantzidis06,
  Vass09}. Therefore, cuspy 
dark matter profiles are robust under major mergers. However, if
baryons are added to dark matter halos, they will presumably condense
more than the dark matter, making up the bulk of the central,
high-density matter in merging halos. As two halos merge, the outer,
less tightly bound and dark-matter-dominated components will be
tidally stripped, but the central high density, predominantly stellar
components will settle into the center of the merger remnant, undergoing
dynamical friction along the way. Therefore, the baryons are an
important ingredient to dry merger scenarios because they ensure the
merging clump is sufficiently tightly bound to reach the central regions of
the nascent galaxy.  

The extreme differences in the inner dark matter halo densities for
the dissipational and dissipationless models emphasize the importance
of the addition of baryons to dark matter halos. Without introducing
modifications to the \LCDM{} paradigm, dark matter halo cusps can be
reduced to cores via the dissipationless formation of the central
stellar regions
of galaxies. The balance between dissipational and dissipationless
formation mechanisms can be probed by observations. Current
observations of BCGs and ellipticals galaxies are sufficient to
exclude a purely dissipational formation mechanism for these 
galaxies. Future measurements of stellar mass-to-light ratios from
microlensing observations, and direct detection of dark matter in the
Milky Way will help to constrain the balance between dissipational and
dissipationless formation mechanisms and the dependence of this
balance on time and environment. 

\acknowledgments We thank S. Tremaine and J. Binney for their valuable
comments and 
corrections. We would also like to thank the referee for his/her
comments and suggestions. CNL acknowledges support from the NDSEG 
fellowship.




%
\bibliographystyle{apj}
\bibliography{lackner}

\begin{thebibliography}{106}
\expandafter\ifx\csname natexlab\endcsname\relax\def\natexlab#1{#1}\fi

\bibitem[{{Abadi} {et~al.}(2009){Abadi}, {Navarro}, {Fardal}, {Babul}, \&
  {Steinmetz}}]{Abadi09}
{Abadi}, M.~G., {Navarro}, J.~F., {Fardal}, M., {Babul}, A., \& {Steinmetz}, M.
  2009, ArXiv e-prints

\bibitem[{{Abadi} {et~al.}(2003){Abadi}, {Navarro}, {Steinmetz}, \&
  {Eke}}]{Abadi03}
{Abadi}, M.~G., {Navarro}, J.~F., {Steinmetz}, M., \& {Eke}, V.~R. 2003, \apj,
  591, 499

\bibitem[{{Aubert} \& {Pichon}(2007)}]{Aubert07}
{Aubert}, D. \& {Pichon}, C. 2007, \mnras, 374, 877

\bibitem[{{Bernardi} {et~al.}(2007){Bernardi}, {Hyde}, {Sheth}, {Miller}, \&
  {Nichol}}]{Bernardi07}
{Bernardi}, M., {Hyde}, J.~B., {Sheth}, R.~K., {Miller}, C.~J., \& {Nichol},
  R.~C. 2007, \aj, 133, 1741

\bibitem[{{Bernardi} {et~al.}(2003){Bernardi}, {Sheth}, {Annis}, {Burles},
  {Eisenstein}, {Finkbeiner}, {Hogg}, {Lupton}, {Schlegel}, {SubbaRao},
  {Bahcall}, {Blakeslee}, {Brinkmann}, {Castander}, {Connolly}, {Csabai},
  {Doi}, {Fukugita}, {Frieman}, {Heckman}, {Hennessy}, {Ivezi{\'c}}, {Knapp},
  {Lamb}, {McKay}, {Munn}, {Nichol}, {Okamura}, {Schneider}, {Thakar}, \&
  {York}}]{Bernardi03I}
{Bernardi}, M., {Sheth}, R.~K., {Annis}, J., {Burles}, S., {Eisenstein}, D.~J.,
  {Finkbeiner}, D.~P., {Hogg}, D.~W., {Lupton}, R.~H., {Schlegel}, D.~J.,
  {SubbaRao}, M., {Bahcall}, N.~A., {Blakeslee}, J.~P., {Brinkmann}, J.,
  {Castander}, F.~J., {Connolly}, A.~J., {Csabai}, I., {Doi}, M., {Fukugita},
  M., {Frieman}, J., {Heckman}, T., {Hennessy}, G.~S., {Ivezi{\'c}}, {\v Z}.,
  {Knapp}, G.~R., {Lamb}, D.~Q., {McKay}, T., {Munn}, J.~A., {Nichol}, R.,
  {Okamura}, S., {Schneider}, D.~P., {Thakar}, A.~R., \& {York}, D.~G. 2003,
  \aj, 125, 1817

\bibitem[{{Bezanson} {et~al.}(2009){Bezanson}, {van Dokkum}, {Tal},
  {Marchesini}, {Kriek}, {Franx}, \& {Coppi}}]{Bezanson09}
{Bezanson}, R., {van Dokkum}, P.~G., {Tal}, T., {Marchesini}, D., {Kriek}, M.,
  {Franx}, M., \& {Coppi}, P. 2009, \apj, 697, 1290

\bibitem[{{Binney} \& {Evans}(2001)}]{Binney01}
{Binney}, J.~J. \& {Evans}, N.~W. 2001, \mnras, 327, L27

\bibitem[{{Blumenthal} {et~al.}(1986){Blumenthal}, {Faber}, {Flores}, \&
  {Primack}}]{Blumenthal86}
{Blumenthal}, G.~R., {Faber}, S.~M., {Flores}, R., \& {Primack}, J.~R. 1986,
  \apj, 301, 27

\bibitem[{{Bode} {et~al.}(2001){Bode}, {Ostriker}, \& {Turok}}]{Bode01}
{Bode}, P., {Ostriker}, J.~P., \& {Turok}, N. 2001, \apj, 556, 93

\bibitem[{{Bode} {et~al.}(2009){Bode}, {Ostriker}, \& {Vikhlinin}}]{Bode09}
{Bode}, P., {Ostriker}, J.~P., \& {Vikhlinin}, A. 2009, \apj, 700, 989

\bibitem[{{Boylan-Kolchin} \& {Ma}(2004)}]{BoylanKolchinMa04}
{Boylan-Kolchin}, M. \& {Ma}, C. 2004, \mnras, 349, 1117

\bibitem[{{Cappellari} {et~al.}(2006){Cappellari}, {Bacon}, {Bureau}, {Damen},
  {Davies}, {de Zeeuw}, {Emsellem}, {Falc{\'o}n-Barroso}, {Krajnovi{\'c}},
  {Kuntschner}, {McDermid}, {Peletier}, {Sarzi}, {van den Bosch}, \& {van de
  Ven}}]{Cappellari06}
{Cappellari}, M., {Bacon}, R., {Bureau}, M., {Damen}, M.~C., {Davies}, R.~L.,
  {de Zeeuw}, P.~T., {Emsellem}, E., {Falc{\'o}n-Barroso}, J., {Krajnovi{\'c}},
  D., {Kuntschner}, H., {McDermid}, R.~M., {Peletier}, R.~F., {Sarzi}, M., {van
  den Bosch}, R.~C.~E., \& {van de Ven}, G. 2006, \mnras, 366, 1126

\bibitem[{{Cappellari} {et~al.}(2009){Cappellari}, {di Serego Alighieri},
  {Cimatti}, {Daddi}, {Renzini}, {Kurk}, {Cassata}, {Dickinson},
  {Franceschini}, {Mignoli}, {Pozzetti}, {Rodighiero}, {Rosati}, \&
  {Zamorani}}]{Cappellari09}
{Cappellari}, M., {di Serego Alighieri}, S., {Cimatti}, A., {Daddi}, E.,
  {Renzini}, A., {Kurk}, J.~D., {Cassata}, P., {Dickinson}, M., {Franceschini},
  A., {Mignoli}, M., {Pozzetti}, L., {Rodighiero}, G., {Rosati}, P., \&
  {Zamorani}, G. 2009, \apjl, 704, L34

\bibitem[{{Colafrancesco} {et~al.}(2006){Colafrancesco}, {Profumo}, \&
  {Ullio}}]{Colafranceso06}
{Colafrancesco}, S., {Profumo}, S., \& {Ullio}, P. 2006, \aap, 455, 21

\bibitem[{{Cole} {et~al.}(2001){Cole}, {Norberg}, {Baugh}, {Frenk},
  {Bland-Hawthorn}, {Bridges}, {Cannon}, {Colless}, {Collins}, {Couch},
  {Cross}, {Dalton}, {De Propris}, {Driver}, {Efstathiou}, {Ellis},
  {Glazebrook}, {Jackson}, {Lahav}, {Lewis}, {Lumsden}, {Maddox}, {Madgwick},
  {Peacock}, {Peterson}, {Sutherland}, \& {Taylor}}]{Cole01}
{Cole}, S., {Norberg}, P., {Baugh}, C.~M., {Frenk}, C.~S., {Bland-Hawthorn},
  J., {Bridges}, T., {Cannon}, R., {Colless}, M., {Collins}, C., {Couch}, W.,
  {Cross}, N., {Dalton}, G., {De Propris}, R., {Driver}, S.~P., {Efstathiou},
  G., {Ellis}, R.~S., {Glazebrook}, K., {Jackson}, C., {Lahav}, O., {Lewis},
  I., {Lumsden}, S., {Maddox}, S., {Madgwick}, D., {Peacock}, J.~A.,
  {Peterson}, B.~A., {Sutherland}, W., \& {Taylor}, K. 2001, \mnras, 326, 255

\bibitem[{{Cook} {et~al.}(2009){Cook}, {Lapi}, \& {Granato}}]{Cook09}
{Cook}, M., {Lapi}, A., \& {Granato}, G.~L. 2009, \mnras, 397, 534

\bibitem[{{Cooray} \& {Milosavljevi{\'c}}(2005)}]{Cooray05}
{Cooray}, A. \& {Milosavljevi{\'c}}, M. 2005, \apjl, 627, L85

\bibitem[{{de Blok} {et~al.}(2008){de Blok}, {Walter}, {Brinks},
  {Trachternach}, {Oh}, \& {Kennicutt}}]{deBlok08}
{de Blok}, W.~J.~G., {Walter}, F., {Brinks}, E., {Trachternach}, C., {Oh},
  S.-H., \& {Kennicutt}, R.~C. 2008, \aj, 136, 2648

\bibitem[{{Dehnen}(2005)}]{Dehnen05}
{Dehnen}, W. 2005, \mnras, 360, 892

\bibitem[{{Desroches} {et~al.}(2007){Desroches}, {Quataert}, {Ma}, \&
  {West}}]{Desroches07}
{Desroches}, L., {Quataert}, E., {Ma}, C., \& {West}, A.~A. 2007, \mnras, 377,
  402

\bibitem[{{Diemand} {et~al.}(2007){Diemand}, {Kuhlen}, \& {Madau}}]{Diemand07}
{Diemand}, J., {Kuhlen}, M., \& {Madau}, P. 2007, \apj, 657, 262

\bibitem[{{Diemand} {et~al.}(2005){Diemand}, {Zemp}, {Moore}, {Stadel}, \&
  {Carollo}}]{Diemand05}
{Diemand}, J., {Zemp}, M., {Moore}, B., {Stadel}, J., \& {Carollo}, C.~M. 2005,
  \mnras, 364, 665

\bibitem[{{Djorgovski} \& {Davis}(1987)}]{DjorgovskiDavis87}
{Djorgovski}, S. \& {Davis}, M. 1987, \apj, 313, 59

\bibitem[{{Dobler} {et~al.}(2007){Dobler}, {Keeton}, \&
  {Wambsganss}}]{Dobler07}
{Dobler}, G., {Keeton}, C.~R., \& {Wambsganss}, J. 2007, \mnras, 377, 977

\bibitem[{{Dressler} {et~al.}(1987){Dressler}, {Lynden-Bell}, {Burstein},
  {Davies}, {Faber}, {Terlevich}, \& {Wegner}}]{Dressler87}
{Dressler}, A., {Lynden-Bell}, D., {Burstein}, D., {Davies}, R.~L., {Faber},
  S.~M., {Terlevich}, R., \& {Wegner}, G. 1987, \apj, 313, 42

\bibitem[{{El-Zant} {et~al.}(2001){El-Zant}, {Shlosman}, \&
  {Hoffman}}]{El-Zant01}
{El-Zant}, A., {Shlosman}, I., \& {Hoffman}, Y. 2001, \apj, 560, 636

\bibitem[{{El-Zant} {et~al.}(2004){El-Zant}, {Hoffman}, {Primack}, {Combes}, \&
  {Shlosman}}]{El-Zant04}
{El-Zant}, A.~A., {Hoffman}, Y., {Primack}, J., {Combes}, F., \& {Shlosman}, I.
  2004, \apjl, 607, L75

\bibitem[{{Faber} \& {Jackson}(1976)}]{FaberJackson76}
{Faber}, S.~M. \& {Jackson}, R.~E. 1976, \apj, 204, 668

\bibitem[{{Flores} \& {Primack}(1994)}]{FloresPrimack94}
{Flores}, R.~A. \& {Primack}, J.~R. 1994, \apjl, 427, L1

\bibitem[{{Fukushige} \& {Makino}(1997)}]{Fukushige97}
{Fukushige}, T. \& {Makino}, J. 1997, \apjl, 477, L9+

\bibitem[{{Gao} {et~al.}(2004){Gao}, {Loeb}, {Peebles}, {White}, \&
  {Jenkins}}]{Gao04}
{Gao}, L., {Loeb}, A., {Peebles}, P.~J.~E., {White}, S.~D.~M., \& {Jenkins}, A.
  2004, \apj, 614, 17

\bibitem[{{Gao} {et~al.}(2008){Gao}, {Navarro}, {Cole}, {Frenk}, {White},
  {Springel}, {Jenkins}, \& {Neto}}]{GaoNavarro08}
{Gao}, L., {Navarro}, J.~F., {Cole}, S., {Frenk}, C.~S., {White}, S.~D.~M.,
  {Springel}, V., {Jenkins}, A., \& {Neto}, A.~F. 2008, \mnras, 387, 536

\bibitem[{{Gavazzi} {et~al.}(2003){Gavazzi}, {Fort}, {Mellier}, {Pell{\'o}}, \&
  {Dantel-Fort}}]{Gavazzi03}
{Gavazzi}, R., {Fort}, B., {Mellier}, Y., {Pell{\'o}}, R., \& {Dantel-Fort}, M.
  2003, \aap, 403, 11

\bibitem[{{Gentile} {et~al.}(2004){Gentile}, {Salucci}, {Klein}, {Vergani}, \&
  {Kalberla}}]{Gentile04}
{Gentile}, G., {Salucci}, P., {Klein}, U., {Vergani}, D., \& {Kalberla}, P.
  2004, \mnras, 351, 903

\bibitem[{{Gentile} {et~al.}(2007){Gentile}, {Tonini}, \&
  {Salucci}}]{Gentile07}
{Gentile}, G., {Tonini}, C., \& {Salucci}, P. 2007, \aap, 467, 925

\bibitem[{{Georgakakis} {et~al.}(2001){Georgakakis}, {Hopkins}, {Caulton},
  {Wiklind}, {Terlevich}, \& {Forbes}}]{Georgakakis01}
{Georgakakis}, A., {Hopkins}, A.~M., {Caulton}, A., {Wiklind}, T., {Terlevich},
  A.~I., \& {Forbes}, D.~A. 2001, \mnras, 326, 1431

\bibitem[{{Gilmore} {et~al.}(2007){Gilmore}, {Wilkinson}, {Wyse}, {Kleyna},
  {Koch}, {Evans}, \& {Grebel}}]{Gilmore07}
{Gilmore}, G., {Wilkinson}, M.~I., {Wyse}, R.~F.~G., {Kleyna}, J.~T., {Koch},
  A., {Evans}, N.~W., \& {Grebel}, E.~K. 2007, \apj, 663, 948

\bibitem[{{Gnedin} {et~al.}(1999){Gnedin}, {Hernquist}, \&
  {Ostriker}}]{Gnedin99}
{Gnedin}, O.~Y., {Hernquist}, L., \& {Ostriker}, J.~P. 1999, \apj, 514, 109

\bibitem[{{Gnedin} {et~al.}(2004){Gnedin}, {Kravtsov}, {Klypin}, \&
  {Nagai}}]{Gnedin04}
{Gnedin}, O.~Y., {Kravtsov}, A.~V., {Klypin}, A.~A., \& {Nagai}, D. 2004, \apj,
  616, 16

\bibitem[{{Gonzalez} {et~al.}(2005){Gonzalez}, {Zabludoff}, \&
  {Zaritsky}}]{Gonzalez05}
{Gonzalez}, A.~H., {Zabludoff}, A.~I., \& {Zaritsky}, D. 2005, \apj, 618, 195

\bibitem[{{Graham} {et~al.}(1996){Graham}, {Lauer}, {Colless}, \&
  {Postman}}]{Graham96}
{Graham}, A., {Lauer}, T.~R., {Colless}, M., \& {Postman}, M. 1996, \apj, 465,
  534

\bibitem[{{Graham}(2007)}]{Graham07}
{Graham}, A.~W. 2007, \mnras, 379, 711

\bibitem[{{Holley-Bockelmann} {et~al.}(2005){Holley-Bockelmann}, {Weinberg}, \&
  {Katz}}]{HolleyBocklemann05}
{Holley-Bockelmann}, K., {Weinberg}, M., \& {Katz}, N. 2005, \mnras, 363, 991

\bibitem[{{Inada} {et~al.}(2008){Inada}, {Oguri}, {Becker}, {Shin}, {Richards},
  {Hennawi}, {White}, {Pindor}, {Strauss}, {Kochanek}, {Johnston}, {Gregg},
  {Kayo}, {Eisenstein}, {Hall}, {Castander}, {Clocchiatti}, {Anderson},
  {Schneider}, {York}, {Lupton}, {Chiu}, {Kawano}, {Scranton}, {Frieman},
  {Keeton}, {Morokuma}, {Rix}, {Turner}, {Burles}, {Brunner}, {Sheldon},
  {Bahcall}, \& {Masataka}}]{Inada08}
{Inada}, N., {Oguri}, M., {Becker}, R.~H., {Shin}, M.-S., {Richards}, G.~T.,
  {Hennawi}, J.~F., {White}, R.~L., {Pindor}, B., {Strauss}, M.~A., {Kochanek},
  C.~S., {Johnston}, D.~E., {Gregg}, M.~D., {Kayo}, I., {Eisenstein}, D.,
  {Hall}, P.~B., {Castander}, F.~J., {Clocchiatti}, A., {Anderson}, S.~F.,
  {Schneider}, D.~P., {York}, D.~G., {Lupton}, R., {Chiu}, K., {Kawano}, Y.,
  {Scranton}, R., {Frieman}, J.~A., {Keeton}, C.~R., {Morokuma}, T., {Rix},
  H.-W., {Turner}, E.~L., {Burles}, S., {Brunner}, R.~J., {Sheldon}, E.~S.,
  {Bahcall}, N.~A., \& {Masataka}, F. 2008, \aj, 135, 496

\bibitem[{{Jardel} \& {Sellwood}(2009)}]{Jardel09}
{Jardel}, J.~R. \& {Sellwood}, J.~A. 2009, \apj, 691, 1300

\bibitem[{{Jesseit} {et~al.}(2002){Jesseit}, {Naab}, \& {Burkert}}]{Jesseit02}
{Jesseit}, R., {Naab}, T., \& {Burkert}, A. 2002, \apjl, 571, L89

\bibitem[{{Johansson} {et~al.}(2009){Johansson}, {Naab}, \&
  {Ostriker}}]{Johansson09}
{Johansson}, P.~H., {Naab}, T., \& {Ostriker}, J.~P. 2009, \apjl, 697, L38

\bibitem[{{Kazantzidis} {et~al.}(2006){Kazantzidis}, {Zentner}, \&
  {Kravtsov}}]{Kazantzidis06}
{Kazantzidis}, S., {Zentner}, A.~R., \& {Kravtsov}, A.~V. 2006, \apj, 641, 647

\bibitem[{{Kormendy}(1977)}]{Kormendy77}
{Kormendy}, J. 1977, \apj, 218, 333

\bibitem[{{Kuhlen} {et~al.}(2008){Kuhlen}, {Diemand}, \& {Madau}}]{Kuhlen08}
{Kuhlen}, M., {Diemand}, J., \& {Madau}, P. 2008, \apj, 686, 262

\bibitem[{{Lauer} {et~al.}(2007){Lauer}, {Faber}, {Richstone}, {Gebhardt},
  {Tremaine}, {Postman}, {Dressler}, {Aller}, {Filippenko}, {Green}, {Ho},
  {Kormendy}, {Magorrian}, \& {Pinkney}}]{Lauer07}
{Lauer}, T.~R., {Faber}, S.~M., {Richstone}, D., {Gebhardt}, K., {Tremaine},
  S., {Postman}, M., {Dressler}, A., {Aller}, M.~C., {Filippenko}, A.~V.,
  {Green}, R., {Ho}, L.~C., {Kormendy}, J., {Magorrian}, J., \& {Pinkney}, J.
  2007, \apj, 662, 808

\bibitem[{{Lima Neto} {et~al.}(1999){Lima Neto}, {Gerbal}, \&
  {M{\'a}rquez}}]{LimaNeto99}
{Lima Neto}, G.~B., {Gerbal}, D., \& {M{\'a}rquez}, I. 1999, \mnras, 309, 481

\bibitem[{{Lin} \& {Mohr}(2004)}]{LinMohr04}
{Lin}, Y.-T. \& {Mohr}, J.~J. 2004, \apj, 617, 879

\bibitem[{{Lin} {et~al.}(2003){Lin}, {Mohr}, \& {Stanford}}]{Lin03}
{Lin}, Y.-T., {Mohr}, J.~J., \& {Stanford}, S.~A. 2003, \apj, 591, 749

\bibitem[{{Loeb} \& {Peebles}(2003)}]{LoebPeebles03}
{Loeb}, A. \& {Peebles}, P.~J.~E. 2003, \apj, 589, 29

\bibitem[{{Ma} \& {Boylan-Kolchin}(2004)}]{MaBoylanKolchin04}
{Ma}, C. \& {Boylan-Kolchin}, M. 2004, Physical Review Letters, 93, 021301

\bibitem[{{McMillan} \& {Dehnen}(2005)}]{McMillanDehnen05}
{McMillan}, P.~J. \& {Dehnen}, W. 2005, \mnras, 363, 1205

\bibitem[{{Milosavljevi{\'c}} \& {Merritt}(2001)}]{Milosavljevic01}
{Milosavljevi{\'c}}, M. \& {Merritt}, D. 2001, \apj, 563, 34

\bibitem[{{Moore} {et~al.}(1999){Moore}, {Quinn}, {Governato}, {Stadel}, \&
  {Lake}}]{Moore99}
{Moore}, B., {Quinn}, T., {Governato}, F., {Stadel}, J., \& {Lake}, G. 1999,
  \mnras, 310, 1147

\bibitem[{{Naab} {et~al.}(2009){Naab}, {Johansson}, \& {Ostriker}}]{Naab09}
{Naab}, T., {Johansson}, P.~H., \& {Ostriker}, J.~P. 2009, \apjl, 699, L178

\bibitem[{{Naab} {et~al.}(2007){Naab}, {Johansson}, {Ostriker}, \&
  {Efstathiou}}]{Naab07}
{Naab}, T., {Johansson}, P.~H., {Ostriker}, J.~P., \& {Efstathiou}, G. 2007,
  \apj, 658, 710

\bibitem[{{Napolitano} {et~al.}(2009){Napolitano}, {Romanowsky}, {Coccato},
  {Capaccioli}, {Douglas}, {Noordermeer}, {Gerhard}, {Arnaboldi}, {de Lorenzi},
  {Kuijken}, {Merrifield}, {O'Sullivan}, {Cortesi}, {Das}, \&
  {Freeman}}]{Napolitano09}
{Napolitano}, N.~R., {Romanowsky}, A.~J., {Coccato}, L., {Capaccioli}, M.,
  {Douglas}, N.~G., {Noordermeer}, E., {Gerhard}, O., {Arnaboldi}, M., {de
  Lorenzi}, F., {Kuijken}, K., {Merrifield}, M.~R., {O'Sullivan}, E.,
  {Cortesi}, A., {Das}, P., \& {Freeman}, K.~C. 2009, \mnras, 393, 329

\bibitem[{{Navarro} \& {Benz}(1991)}]{Navarro91}
{Navarro}, J.~F. \& {Benz}, W. 1991, \apj, 380, 320

\bibitem[{{Navarro} {et~al.}(1997){Navarro}, {Frenk}, \& {White}}]{NFW97}
{Navarro}, J.~F., {Frenk}, C.~S., \& {White}, S.~D.~M. 1997, \apj, 490, 493

\bibitem[{{Navarro} \& {White}(1994)}]{Navarro94}
{Navarro}, J.~F. \& {White}, S.~D.~M. 1994, \mnras, 267, 401

\bibitem[{{Neto} {et~al.}(2007){Neto}, {Gao}, {Bett}, {Cole}, {Navarro},
  {Frenk}, {White}, {Springel}, \& {Jenkins}}]{Neto07}
{Neto}, A.~F., {Gao}, L., {Bett}, P., {Cole}, S., {Navarro}, J.~F., {Frenk},
  C.~S., {White}, S.~D.~M., {Springel}, V., \& {Jenkins}, A. 2007, \mnras, 381,
  1450

\bibitem[{{Nipoti} {et~al.}(2004){Nipoti}, {Treu}, {Ciotti}, \&
  {Stiavelli}}]{Nipoti04}
{Nipoti}, C., {Treu}, T., {Ciotti}, L., \& {Stiavelli}, M. 2004, \mnras, 355,
  1119

\bibitem[{{Oh} {et~al.}(2008){Oh}, {de Blok}, {Walter}, {Brinks}, \&
  {Kennicutt}}]{Oh08}
{Oh}, S.-H., {de Blok}, W.~J.~G., {Walter}, F., {Brinks}, E., \& {Kennicutt},
  R.~C. 2008, \aj, 136, 2761

\bibitem[{{Ostriker} \& {Hausman}(1977)}]{OstrikerHausman77}
{Ostriker}, J.~P. \& {Hausman}, M.~A. 1977, \apjl, 217, L125

\bibitem[{{Paczynski}(1986)}]{Paczynski86}
{Paczynski}, B. 1986, \apj, 301, 503

\bibitem[{{Pahre} {et~al.}(1998){Pahre}, {de Carvalho}, \&
  {Djorgovski}}]{Pahre98}
{Pahre}, M.~A., {de Carvalho}, R.~R., \& {Djorgovski}, S.~G. 1998, \aj, 116,
  1606

\bibitem[{{Paturel} {et~al.}(2003){Paturel}, {Petit}, {Prugniel}, {Theureau},
  {Rousseau}, {Brouty}, {Dubois}, \& {Cambr{\'e}sy}}]{Paturel03}
{Paturel}, G., {Petit}, C., {Prugniel}, P., {Theureau}, G., {Rousseau}, J.,
  {Brouty}, M., {Dubois}, P., \& {Cambr{\'e}sy}, L. 2003, \aap, 412, 45

\bibitem[{{Peirani} {et~al.}(2008){Peirani}, {Kay}, \& {Silk}}]{Peirani08}
{Peirani}, S., {Kay}, S., \& {Silk}, J. 2008, \aap, 479, 123

\bibitem[{{Pichon} \& {Aubert}(2006)}]{Aubert06}
{Pichon}, C. \& {Aubert}, D. 2006, \mnras, 368, 1657

\bibitem[{{Pooley} {et~al.}(2009){Pooley}, {Rappaport}, {Blackburne},
  {Schechter}, {Schwab}, \& {Wambsganss}}]{Pooley09}
{Pooley}, D., {Rappaport}, S., {Blackburne}, J., {Schechter}, P.~L., {Schwab},
  J., \& {Wambsganss}, J. 2009, \apj, 697, 1892

\bibitem[{{Postman} \& {Lauer}(1995)}]{Postman95}
{Postman}, M. \& {Lauer}, T.~R. 1995, \apj, 440, 28

\bibitem[{{Rhee} {et~al.}(2004){Rhee}, {Valenzuela}, {Klypin}, {Holtzman}, \&
  {Moorthy}}]{Rhee04}
{Rhee}, G., {Valenzuela}, O., {Klypin}, A., {Holtzman}, J., \& {Moorthy}, B.
  2004, \apj, 617, 1059

\bibitem[{{Ricotti}(2003)}]{Ricotti03}
{Ricotti}, M. 2003, \mnras, 344, 1237

\bibitem[{{Romano-D{\'{\i}}az} {et~al.}(2009){Romano-D{\'{\i}}az}, {Shlosman},
  {Heller}, \& {Hoffman}}]{RomanoDiaz09}
{Romano-D{\'{\i}}az}, E., {Shlosman}, I., {Heller}, C., \& {Hoffman}, Y. 2009,
  ArXiv e-prints

\bibitem[{{Romano-D{\'{\i}}az} {et~al.}(2008){Romano-D{\'{\i}}az}, {Shlosman},
  {Hoffman}, \& {Heller}}]{Romano-Diaz08}
{Romano-D{\'{\i}}az}, E., {Shlosman}, I., {Hoffman}, Y., \& {Heller}, C. 2008,
  \apjl, 685, L105

\bibitem[{{Romanowsky} {et~al.}(2003){Romanowsky}, {Douglas}, {Arnaboldi},
  {Kuijken}, {Merrifield}, {Napolitano}, {Capaccioli}, \&
  {Freeman}}]{Romanowsky03}
{Romanowsky}, A.~J., {Douglas}, N.~G., {Arnaboldi}, M., {Kuijken}, K.,
  {Merrifield}, M.~R., {Napolitano}, N.~R., {Capaccioli}, M., \& {Freeman},
  K.~C. 2003, Science, 301, 1696

\bibitem[{{Sand} {et~al.}(2004){Sand}, {Treu}, {Smith}, \&
  {Ellis}}]{SandTreu04}
{Sand}, D.~J., {Treu}, T., {Smith}, G.~P., \& {Ellis}, R.~S. 2004, \apj, 604,
  88

\bibitem[{{Schechter} \& {Wambsganss}(2002)}]{SchechterWambsganss02}
{Schechter}, P.~L. \& {Wambsganss}, J. 2002, \apj, 580, 685

\bibitem[{{Sersic}(1968)}]{Sersic69}
{Sersic}, J.~L. 1968, {Atlas de galaxias australes} (Cordoba, Argentina:
  Observatorio Astronomico, 1968)

\bibitem[{{Simon} {et~al.}(2005){Simon}, {Bolatto}, {Leroy}, {Blitz}, \&
  {Gates}}]{Simon05}
{Simon}, J.~D., {Bolatto}, A.~D., {Leroy}, A., {Blitz}, L., \& {Gates}, E.~L.
  2005, \apj, 621, 757

\bibitem[{{Spekkens} {et~al.}(2005){Spekkens}, {Giovanelli}, \&
  {Haynes}}]{Spekkens05}
{Spekkens}, K., {Giovanelli}, R., \& {Haynes}, M.~P. 2005, \aj, 129, 2119

\bibitem[{{Spergel} \& {Steinhardt}(2000)}]{SpergelSteinhardt00}
{Spergel}, D.~N. \& {Steinhardt}, P.~J. 2000, Physical Review Letters, 84, 3760

\bibitem[{{Springel} {et~al.}(2008){Springel}, {Wang}, {Vogelsberger},
  {Ludlow}, {Jenkins}, {Helmi}, {Navarro}, {Frenk}, \& {White}}]{Springel08}
{Springel}, V., {Wang}, J., {Vogelsberger}, M., {Ludlow}, A., {Jenkins}, A.,
  {Helmi}, A., {Navarro}, J.~F., {Frenk}, C.~S., \& {White}, S.~D.~M. 2008,
  \mnras, 391, 1685

\bibitem[{{Stoehr} {et~al.}(2003){Stoehr}, {White}, {Springel}, {Tormen}, \&
  {Yoshida}}]{Stoehr03}
{Stoehr}, F., {White}, S.~D.~M., {Springel}, V., {Tormen}, G., \& {Yoshida}, N.
  2003, \mnras, 345, 1313

\bibitem[{{Strigari} {et~al.}(2007){Strigari}, {Koushiappas}, {Bullock}, \&
  {Kaplinghat}}]{Strigari07}
{Strigari}, L.~E., {Koushiappas}, S.~M., {Bullock}, J.~S., \& {Kaplinghat}, M.
  2007, \prd, 75, 083526

\bibitem[{{Subramanian} {et~al.}(2000){Subramanian}, {Cen}, \&
  {Ostriker}}]{Subramanian00}
{Subramanian}, K., {Cen}, R., \& {Ostriker}, J.~P. 2000, \apj, 538, 528

\bibitem[{{Swaters} {et~al.}(2003){Swaters}, {Madore}, {van den Bosch}, \&
  {Balcells}}]{Swaters03}
{Swaters}, R.~A., {Madore}, B.~F., {van den Bosch}, F.~C., \& {Balcells}, M.
  2003, \apj, 583, 732

\bibitem[{{Tonini} {et~al.}(2006){Tonini}, {Lapi}, \& {Salucci}}]{Tonini06}
{Tonini}, C., {Lapi}, A., \& {Salucci}, P. 2006, \apj, 649, 591

\bibitem[{{Tremaine} \& {Weinberg}(1984)}]{TremaineWeinberg84}
{Tremaine}, S. \& {Weinberg}, M.~D. 1984, \mnras, 209, 729

\bibitem[{{Treu} {et~al.}(2006){Treu}, {Koopmans}, {Bolton}, {Burles}, \&
  {Moustakas}}]{Treu06}
{Treu}, T., {Koopmans}, L.~V., {Bolton}, A.~S., {Burles}, S., \& {Moustakas},
  L.~A. 2006, \apj, 640, 662

\bibitem[{{Treu} \& {Koopmans}(2004)}]{TreuKoopmans04}
{Treu}, T. \& {Koopmans}, L.~V.~E. 2004, \apj, 611, 739

\bibitem[{{Vale} \& {Ostriker}(2008)}]{ValeOstriker08}
{Vale}, A. \& {Ostriker}, J.~P. 2008, \mnras, 383, 355

\bibitem[{{Valenzuela} {et~al.}(2007){Valenzuela}, {Rhee}, {Klypin},
  {Governato}, {Stinson}, {Quinn}, \& {Wadsley}}]{Valenzuela07}
{Valenzuela}, O., {Rhee}, G., {Klypin}, A., {Governato}, F., {Stinson}, G.,
  {Quinn}, T., \& {Wadsley}, J. 2007, \apj, 657, 773

\bibitem[{{van der Wel} {et~al.}(2004){van der Wel}, {Franx}, {van Dokkum}, \&
  {Rix}}]{vanderWel04}
{van der Wel}, A., {Franx}, M., {van Dokkum}, P.~G., \& {Rix}, H.-W. 2004,
  \apjl, 601, L5

\bibitem[{{van Dokkum} {et~al.}(2008){van Dokkum}, {Franx}, {Kriek}, {Holden},
  {Illingworth}, {Magee}, {Bouwens}, {Marchesini}, {Quadri}, {Rudnick},
  {Taylor}, \& {Toft}}]{vanDokkum08}
{van Dokkum}, P.~G., {Franx}, M., {Kriek}, M., {Holden}, B., {Illingworth},
  G.~D., {Magee}, D., {Bouwens}, R., {Marchesini}, D., {Quadri}, R., {Rudnick},
  G., {Taylor}, E.~N., \& {Toft}, S. 2008, \apjl, 677, L5

\bibitem[{{Vass} {et~al.}(2009){Vass}, {Kazantzidis}, {Valluri}, \&
  {Kravtsov}}]{Vass09}
{Vass}, I.~M., {Kazantzidis}, S., {Valluri}, M., \& {Kravtsov}, A.~V. 2009,
  \apj, 698, 1813

\bibitem[{{Wambsganss}(2006)}]{Wambsganss06}
{Wambsganss}, J. 2006, {Gravitational Microlensing} (Gravitational Lensing:
  Strong, Weak and Micro, Saas-Fee Advanced Courses, Volume 33.~ISBN
  978-3-540-30309-1.~Springer-Verlag Berlin Heidelberg, 2006, p.~453), 453--+

\bibitem[{{Weinberg}(1986)}]{Weinberg86}
{Weinberg}, M.~D. 1986, \apj, 300, 93

\bibitem[{{Weinberg} \& {Katz}(2002)}]{Weinberg02}
{Weinberg}, M.~D. \& {Katz}, N. 2002, \apj, 580, 627

\bibitem[{{Wetzel} \& {White}(2009)}]{Wetzel09}
{Wetzel}, A.~R. \& {White}, M. 2009, ArXiv e-prints

\bibitem[{{Wo{\'z}niak} {et~al.}(2000){Wo{\'z}niak}, {Udalski},
  {Szyma{\'n}ski}, {Kubiak}, {Pietrzy{\'n}ski}, {Soszy{\'n}ski}, \&
  {{\.Z}ebru{\'n}}}]{Wozniak00}
{Wo{\'z}niak}, P.~R., {Udalski}, A., {Szyma{\'n}ski}, M., {Kubiak}, M.,
  {Pietrzy{\'n}ski}, G., {Soszy{\'n}ski}, I., \& {{\.Z}ebru{\'n}}, K. 2000,
  \apjl, 540, L65

\end{thebibliography}

\tailfig{\figcartoon}
\tailfig{\figvcirc}
\tailfig{\figLsigma}
\tailfig{\figlenses}
\tailfig{\figmsbestfit}
\tailfig{\figMratio}
\tailfig{\figSAURON}
\tailfig{\figPNe}
\tailfig{\figPNeDMdensity}
\tailfig{\tablemodels}

\end{document}